\documentclass[11pt]{article}
\usepackage{jheppub}
\usepackage{graphicx}
\usepackage{amsmath,amssymb}
\usepackage{array}
\usepackage{comment}
\usepackage{hyperref}
\usepackage{pb-diagram}
\usepackage{slashed}
\usepackage{cancel}
\usepackage{float}
\usepackage[normalem]{ulem}
\usepackage{tcolorbox}
\usepackage{physics}
\usepackage{listings} 
\usepackage{tikz}
\usetikzlibrary{shapes.geometric, arrows}
\tikzstyle{decision} = [ellipse, minimum width=3cm, minimum height=1cm, text centered, draw=black, text width=2cm, fill=green!30]
\tikzstyle{rect} = [rectangle, rounded corners, minimum width=1cm, minimum height=1cm,text centered, fill=yellow, draw=black]
\tikzstyle{trap} = [trapezium, rounded corners,minimum width=1cm, centered, fill=orange, draw=black]
\tikzstyle{arrow} = [thick,->,>=stealth]

\newcommand{\be}{\begin{equation}}
\newcommand{\ee}{\end{equation}}
\newcommand{\bea}{\begin{eqnarray}}
\newcommand{\eea}{\end{eqnarray}}

\newcommand{\doublet}[2]{ \left( \begin{array}{c}#1 \\ [1ex] #2 \end{array}\right) }

\newcommand{\Sqhalf}{{\textstyle\frac{1}{\sqrt{2}}}}
\newcommand{\VSqhalf}[1]{{\textstyle\frac{ #1 }{\sqrt{2}}}}

\newcommand{\Z}{\mathbb{Z}}
\newcommand{\G}{\mathbb{G}}

\newcommand{\GeV}{\mathrm{\;GeV}}
\newcommand{\Veff}{V_{\text{eff}} }

\graphicspath{{Figures/}}

\usepackage{color}
\usepackage{xcolor}
\definecolor{MyRed}{rgb}{0.9,0.1,0.1}

\definecolor{MyBlue}{rgb}{0.1,0.4,0.8}

\definecolor{MyGreen}{cmyk}{0.76,0,0.76,0.45}

\definecolor{MyBrown}{rgb}{0.396,0.263,0.129}

\definecolor{MyGrey}{cmyk}{0,0,0,0.75}

\definecolor{MyOrange}{cmyk}{0,0.5,1,0}

\definecolor{MyYellow}{cmyk}{0,0,1,0.2}

\definecolor{MyCyan}{cmyk}{1,0,0,0.2}

\definecolor{MyMagenta}{cmyk}{0,1,0,0.2}

\definecolor{Jpurple}{rgb}{255,0,255}

\begin{document} 

\title{\hfill ~ \\[-40mm]
\begin{footnotesize}
\hspace{126mm}
\normalfont{DIAS-STP-25-13}\\
\end{footnotesize}
\vspace{30mm}
\boldmath Electroweak phase transition enhanced by a CP-violating dark sector }
\author[\,a,b]{Venus Keus,}
\author[\,c]{Lucy Lewitt,}
\author[\,a]{Jasmine Thomson-Cooke}

\affiliation[a]{School of Theoretical Physics, Dublin Institute for Advanced Studies, 10 Burlington Road, Dublin, D04 C932, Ireland}
\affiliation[b]{Department of Physics and Helsinki Institute of Physics, Gustaf Hallstromin katu 2, Helsinki, FIN-00014, University of Helsinki, Finland}
\affiliation[c]{Department of Maths and Physical Sciences, University of Sheffield,  Western Bank, Sheffield, S10 2TN, United Kingdom}

\emailAdd{venus@stp.dias.ie}
\emailAdd{llewitt@cern.ch}
\emailAdd{jasmine@stp.dias.ie}
\abstract{
Within a well-motivated 3-Higgs doublet model, in which the extended dark sector accommodates CP violation, we analyse the electroweak phase transition (EWPT) at one- and two-loop order. We show the importance of higher loop calculations in EWPT analyses and identify the regions of the parameter space of our model where EWPT is of first order while in agreement with all theoretical and experimental bounds, including Dark Matter relic density and direct and indirect searches.
\\ \\ \\ \today
}

\maketitle

\section{Introduction}
\label{sec:intro}

The Standard Model (SM) of particle physics has undergone extensive experimental verification and remains in remarkable agreement with observations. The discovery of its last missing component, the Higgs boson, at the LHC in 2012~\cite{Aad:2012tfa,Chatrchyan:2012ufa} solidified its predictive power. Despite ongoing searches, no substantial deviations from the SM have been observed at the LHC, and the properties of the discovered scalar particle align well with those of the SM Higgs boson~\cite{Flechl:2019jnr,Aad:2019mbh}.

However, the SM does not account for several fundamental phenomena, including the observed baryon asymmetry of the Universe and the existence of a viable Dark Matter (DM) candidate. Various astrophysical observations suggest the necessity of a DM particle that is cosmologically stable, cold, non-baryonic, electrically neutral, and weakly interacting. No such particle exists within the SM framework. Furthermore, the amount of CP violation predicted by the SM is several orders of magnitude smaller than what is required to generate the observed matter-antimatter asymmetry~\cite{Gavela:1993ts,Huet:1994jb,Gavela:1994dt}.

These shortcomings strongly indicate the need for beyond Standard Model (BSM) physics in the search for a more complete theory of Nature. Among the simplest BSM extensions addressing the issues mentioned above, are non-minimal Higgs sectors, predicting the Higgs boson discovered at the LHC is part of a broader scalar sector. Extending the scalar sector can introduce new sources of CP violation and provide viable DM candidates. The scalar potential remains one of the least constrained sectors of the SM, making it a promising avenue for such extensions. Additionally, non-minimal Higgs sectors incorporating discrete symmetries naturally, allowing for Weakly Interacting Massive Particles (WIMPs)~\cite{Jungman:1995df,Bertone:2004pz,Bergstrom:2000pn,Ivanov:2012hc}, which are among the most widely studied DM candidates. The stability of these WIMPs is ensured by conserved discrete symmetries, allowing them to achieve a relic abundance consistent with observation~\cite{Ade:2015xua} via the freeze-out mechanism. 

Extensive research has been conducted on one-singlet and one-doublet scalar extensions of the SM, particularly in the form of Higgs portal models and 2-Higgs-doublet models (2HDMs) (see, e.g.,~\cite{Englert:2011yb,Branco:2011iw} and references therein). However, these models, by design, offer only partial solutions to the unresolved problems of the SM;
notably, the well-established $\Z_2$-symmetric Higgs portal model~\cite{Bertolami:2007wb} and the Inert Doublet Model (IDM)~\cite{Deshpande:1977rw}, are consistent with both direct and indirect DM search constraints. However, they face stringent limitations from LHC data, particularly from measurements of the Higgs invisible decay width and Higgs signal strength. Moreover, the scalar potential in these models is inherently CP conserving.
If one disregards the requirement of a DM candidate, the 2HDM scalar potential can, in principle, accommodate CP violation. However, the introduction of new CP-violating interactions modifies SM Higgs couplings and contributes to the electric dipole moments (EDMs) of the neutron, electron, and certain atomic nuclei~\cite{Chupp:2017rkp}, leading to very strong experimental constraints~\cite{Inoue:2014nva,Keus:2015hva,Keus:2017ioh,Yamanaka:2017mef}. 
On the other hand, purely singlet scalar extensions of the SM remain CP conserving regardless of apparent phases in the vevs or potential parameters.\footnote{Due to the absence of gauge interactions, there is significant arbitrariness in defining charge conjugation for singlet scalars, making these models inherently CP conserving~\cite{Branco:1999fs}.} This points to the fact that for the scalar sector to yield CP violation and a viable DM candidate, inevitably, it needs to be extended beyond the simple one singlet or one doublet setup. 

In contrast to the simple scenarios mentioned before, more elaborate scalar extensions, such as 3-Higgs-doublet models (3HDMs), provide viable DM candidates, introduce new sources of CP violation, enable a strongly first-order phase transition necessary for baryogenesis, incorporate inflaton candidates relevant for the early universe inflationary epoch, and address the fermion mass hierarchy problem, all in a single framework. These features arise due to various symmetries and symmetry-breaking patterns within the scalar potential, which dictate the number of \textit{active} (developing a vacuum expectation value (vev)) and \textit{inert} (lacking a vev) scalar doublets~\cite{Weinberg:1976hu,Ivanov:2012fp,Keus:2014jha,Keus:2014isa,Keus:2015xya,Cordero-Cid:2016krd,Cordero:2017owj,Cordero-Cid:2018man,Keus:2019szx,Cordero-Cid:2020yba,Keus:2021dti,Keus:2013hya,Hernandez-Sanchez:2020aop,Hernandez-Sanchez:2022dnn,Dey:2024epo,Keus:2024khd,Dey:2023exa}.
If CP violation is introduced in the active scalar sector while a separate inert sector provides the DM candidate, the model faces similar constraints to those found in 2HDMs, IDMs, and Higgs portal models~\cite{Grzadkowski:2009bt,Osland:2013sla}. However, extending the inert sector to host both DM and CP violation alleviates these constraints.

The concept of dark or inert CP violation was first introduced in~\cite{Cordero-Cid:2016krd} and further developed in~\cite{Keus:2016orl,Keus:2019szx,Cordero:2017owj,Cordero-Cid:2018man,Cordero-Cid:2020yba} within the framework of a 3HDM. In this scenario, CP-mixed dark scalars interact with the SM particles via Higgs and gauge boson interactions. These studies have demonstrated that dark-sector CP violation allows for more flexibility in gauge-scalar couplings and significantly expands the viable parameter space for a CP-violating DM candidate that aligns with cosmological and collider constraints. Additionally, since the inert sector does not directly couple to the SM fermions, its CP-violating effects evade the stringent bounds imposed by EDM measurements.\footnote{It is important to note that if the extended inert sector consists of a singlet and a doublet rather than two doublets, the available CP violation is inherently reduced due to the presence of the singlet~\cite{Azevedo:2018fmj}. Moreover, the number of possible co-annihilation channels for the DM candidate is smaller, leading to a more constrained and less favourable model compared to the 3HDM.}
As a result, one can construct a CP-violating DM model with unbounded dark CP violation. In fact, the CP-violating dark particles need not have a Higgs-DM coupling and can interact with the SM merely through the gauge bosons, ridding the model of all current (in)direct detection and LHC bounds, while yielding relic abundance in agreement with observation through the freeze-out or freeze-in mechanisms~\cite{Keus:2019szx}.

Given the success of 3HDMs as minimal BSM frameworks accommodating both CP violation and DM, it is natural to ask whether these models can also support a strong first-order electroweak phase transition (EWPT), as required by the mechanism of electroweak baryogenesis~\cite{Kuzmin:1985mm,Cohen:1990it,Turok:1990in,Morrissey:2012db}. This mechanism is particularly appealing due to its testability, in contrast to alternatives such as leptogenesis or GUT- and Planck-scale baryogenesis, which may remain experimentally inaccessible (see~\cite{Riotto:1998bt} for a comprehensive review).
Electroweak baryogenesis assumes no initial baryon asymmetry in the symmetric phase of the early universe. As the temperature drops below the electroweak scale ($ \sim 100$ GeV), the Higgs field develops a non-zero vev, spontaneously breaking electroweak symmetry. If the resulting phase transition is first order, bubbles of the broken phase nucleate within the symmetric phase and coalesce until they fill the entire space. The strength of the EWPT is thus a key component in the viability of the electroweak baryogenesis mechanism in a given model.

In the SM, perturbative analyses of the finite-temperature effective potential indicate a smooth crossover for the observed Higgs mass of $125$ GeV~\cite{Arnold:1992rz,Kajantie:1996qd}. Non-perturbative lattice studies confirm this, ruling out a first-order EWPT for Higgs masses above 80 GeV~\cite{Kajantie:1996mn,Rummukainen:1998as,Laine:1998qk}.
Extensions of the SM with additional scalar fields modify the Higgs potential, potentially strengthening the EWPT at finite temperature. Such studies typically rely on perturbative methods, as non-perturbative simulations are computationally expensive. 
However, perturbative approaches to finite-temperature field theory are hindered by infrared divergences, gauge and renormalisation ambiguities, slow convergence, and difficulties in computing quantities such as the surface tension, which critically impact the determination of transition strength~\cite{Patel:2011th,Garny:2012cg,DOnofrio:2014rug}. 
For instance, in the SM, perturbative methods significantly underestimated the transition temperature compared to non-perturbative results~\cite{Kajantie:1995kf,Arnold:1992rz,Kajantie:1996qd}.
Perturbative approaches in 2HDMs and singlet-extended models have naively demonstrated the possibility of achieving a strong first-order transition, provided the additional scalars introduce significant modifications to the Higgs thermal potential~\cite{Fromme:2006cm,Profumo:2007wc,Espinosa:2011ax,Cline:2012hg,Beniwal:2017eik,Carena:2018vpt,Alanne:2018zjm,Alanne:2020jwx}. However, non-perturbative lattice simulations have shown that such transitions, while possible, are typically confined to narrow and often fine-tuned regions of parameter space, and that perturbative analyses tend to significantly overestimate both the strength and the critical temperature of the transition~\cite{Kainulainen:2019kyp,Niemi:2021qvp}.

In more elaborate scenarios such as 3HDMs and multi-scalar extensions, naive one-loop perturbative studies indicate an increased likelihood of strong first-order phase transitions~\cite{Ahriche:2015mea,Basler:2024aaf,Biermann:2022meg}. However, owing to computational challenges, systematic perturbative analyses and non-perturbative studies of the EWPT in these models remain limited, leaving their phase structure poorly characterised. We aim to bridge this gap by developing precise perturbative tools that can later be extended to non-perturbative lattice simulations.

Given the computational difficulty of fully non-perturbative treatments, it is highly advantageous to first constrain the viable parameter space using rigorous perturbative tools. As a first step in this direction, we compute the EWPT at two-loop level in a well-motivated 3HDM featuring a CP-violating dark sector.
Our analysis follows the methodology of existing lattice simulations~\cite{Brauner:2016fla,Andersen:2017ika,Gorda:2018hvi,Niemi:2018asa,Kainulainen:2019kyp} which are typically based on dimensionally reduced high-temperature effective theories. These effective theories, constructed at two-loop order by integrating out heavy modes, improve upon one-loop analyses and simplify both perturbative and non-perturbative treatments. In many cases, they retain the SM field content, enabling direct use of established lattice frameworks.

The remainder of the paper is organised as follows. Section~\ref{sec:potential} introduces the model and parameter inputs. Theoretical and experimental constraints, together with the chosen benchmark scenarios, are discussed in Section~\ref{sec:constraints}. Section~\ref{sec:thermal} outlines the thermal corrections and construction of the effective potential while Section~\ref{sec:EWPT} presents our numerical setup and results. We conclude and discuss future directions in Section~\ref{sec:conclusions}.

\section{The extended scalar sector}
\label{sec:potential}
In models with $N$-Higgs doublets, the scalar potential invariant under a discrete group \( \G \) of phase rotations can be written as the sum of two parts: a generic \( V_0 \) part, invariant under global continuous phase rotations, and a symmetry-specific \( V_\G \) part, invariant only under the discrete group \( \G \)~\cite{Ivanov:2011ae,Keus:2013hya}. Accordingly, one can write the scalar potential for a \( \Z_2 \)-symmetric 3HDM, as follows:
\begin{eqnarray}
\label{eq:V0-3HDM}
V_{\mathrm{3HDM}}&=&V_0+V_{\Z_2}, \\
V_0 &=& - \mu^2_{1} (\phi_1^\dagger \phi_1) -\mu^2_2 (\phi_2^\dagger \phi_2) - \mu^2_3(\phi_3^\dagger \phi_3) \nonumber\\
&&+ \lambda_{11} (\phi_1^\dagger \phi_1)^2+ \lambda_{22} (\phi_2^\dagger \phi_2)^2  + \lambda_{33} (\phi_3^\dagger \phi_3)^2 \nonumber\\
&& + \lambda_{12}  (\phi_1^\dagger \phi_1)(\phi_2^\dagger \phi_2)
 + \lambda_{23}  (\phi_2^\dagger \phi_2)(\phi_3^\dagger \phi_3) + \lambda_{31} (\phi_3^\dagger \phi_3)(\phi_1^\dagger \phi_1) \nonumber\\
&& + \lambda'_{12} (\phi_1^\dagger \phi_2)(\phi_2^\dagger \phi_1) 
 + \lambda'_{23} (\phi_2^\dagger \phi_3)(\phi_3^\dagger \phi_2) + \lambda'_{31} (\phi_3^\dagger \phi_1)(\phi_1^\dagger \phi_3),  \nonumber\\
 V_{\Z_2} &=& -\mu^2_{12}(\phi_1^\dagger\phi_2)+  \lambda_{1}(\phi_1^\dagger\phi_2)^2 + \lambda_2(\phi_2^\dagger\phi_3)^2 + \lambda_3(\phi_3^\dagger\phi_1)^2  + h.c., \nonumber
\end{eqnarray}
where the three Higgs doublets, $\phi_1,\phi_2$ and $\phi_3$, transform under the $\Z_2$ group, respectively, as 
\begin{equation}
\label{eq:generator}
g_{\Z_2}=  \mathrm{\rm diag}\left(-1, -1, +1 \right). 
\end{equation}

The four explicitly \( \Z_2 \)-symmetric terms included in \( V_{\Z_2} \) ensure that the scalar potential of the model is invariant only under the \( \Z_2 \) group without any accidental enhancement to a larger symmetry~\cite{Ivanov:2011ae}. While additional \( \Z_2 \)-invariant operators, such as 
\( (\phi_1^\dagger\phi_2)(\phi_1^\dagger\phi_1) \), 
\( (\phi_1^\dagger\phi_2)(\phi_2^\dagger\phi_2) \), 
\( (\phi_3^\dagger\phi_1)(\phi_2^\dagger\phi_3) \), and 
\( (\phi_1^\dagger\phi_2)(\phi_3^\dagger\phi_3) \), are allowed by symmetry, they do not qualitatively affect the phenomenology of the model.\footnote{The terms $ 
(\phi_1^\dagger\phi_2)(\phi_1^\dagger\phi_1), \,(\phi_1^\dagger\phi_2)(\phi_2^\dagger\phi_2)$ appear only in inert scalar self-interactions and are irrelevant to our analysis. The terms $(\phi_3^\dagger\phi_1)(\phi_2^\dagger\phi_3), \,(\phi_1^\dagger\phi_2)(\phi_3^\dagger\phi_3)$ effectively shift the values of the $\lambda_{2,3}$ parameters. Our numerical study confirms that excluding these terms does not reduce the generality of the model.} To simplify the analysis, we, therefore, omit these operators by setting their coefficients to zero.

The three scalar doublets are defined as
\begin{equation}
\phi_1 = \doublet{H^+_1}{\Sqhalf (H_1 + i \, A_1)}, \quad
\phi_2 = \doublet{H^+_2}{\Sqhalf (H_2 + i \, A_2)}, \quad
\phi_3 = \doublet{G^+}{\Sqhalf (v + h + i \, G^0)}, 
\label{eq:explicit-fields}
\end{equation}
where \( \phi_1 \) and \( \phi_2 \) are the \( \Z_2 \)-odd \textit{inert} doublets with vanishing vevs, \( \langle \phi_1 \rangle = \langle \phi_2 \rangle = 0 \), while \( \phi_3 \) is the \( \Z_2 \)-even \textit{active} doublet developing a non-zero vev, \( \langle \phi_3 \rangle = \VSqhalf{v}\). The field \( h \) is identified with the physical Higgs boson of the SM, while \( G^\pm \) and \( G^0 \) are the corresponding would-be Goldstone bosons.

The \( \Z_2 \)-charge assignment follows the generator defined in Eq.~\eqref{eq:generator}, with the inert doublets \( \phi_1 \) and \( \phi_2 \) assigned odd parity and the active doublet \( \phi_3 \) assigned even parity. Consequently, the vacuum configuration \( (0, 0, \VSqhalf{v} )\) respects the \( \Z_2 \) symmetry of the scalar potential.

The CP-even scalar \( h \), originating from the active doublet \( \phi_3 \), has tree-level couplings identical to those of the SM Higgs boson. As a result, all CP-violating effects are confined to the inert sector, which remains decoupled from the active sector due to the exact \( \Z_2 \) symmetry. 
This separation permits arbitrarily large CP violation in the inert sector without violating experimental bounds on EDMs. This phenomena, known as \textit{dark CP violation}, was first proposed in~\cite{Cordero-Cid:2016krd}. The DM candidate in this model is the lightest neutral state arising from CP mixing in the inert doublets,\footnote{We consider only regions of parameter space where the lightest inert state is neutral, excluding scenarios in which a charged inert scalar is the lightest.} and its stability is ensured by the unbroken \( \Z_2 \) symmetry.

As noted earlier, the \( \Z_2 \) symmetry  is extended to the full Lagrangian by assigning even \( \Z_2 \)-parity to all SM gauge bosons and fermions. The Yukawa sector is then implemented in a Type-I configuration, in which only the third doublet \( \phi_3 \) couples to fermions:
\begin{equation}
\mathcal{L}_{Y} = \Gamma^u_{mn} \, \bar{q}_{m,L} \, \tilde{\phi}_3 \, u_{n,R} 
+ \Gamma^d_{mn} \, \bar{q}_{m,L} \, \phi_3 \, d_{n,R}
+ \Gamma^e_{mn} \, \bar{l}_{m,L} \, \phi_3 \, e_{n,R}
+ \Gamma^{\nu}_{mn} \, \bar{l}_{m,L} \, \tilde{\phi}_3 \, \nu_{n,R} + \text{h.c.}
\label{eq:yukawa}
\end{equation}
This structure eliminates tree-level flavour-changing neutral currents (FCNCs). Moreover, since \( \phi_1 \) and \( \phi_2 \) do not acquire vevs, they remain completely decoupled from the fermionic sector.


\subsection{Explicit CP-violation}
\label{sec:CPV-details}
The parameters of the phase-invariant part of the potential, $V_0$, are real by construction. Explicit CP violation is introduced via complex parameters $\mu^2_{12}, \lambda_1,\lambda_2, \lambda_3$ in the potential defined in Eq.~\eqref{eq:V0-3HDM}.
The following notation will be employed consistently throughout this paper:
\be
\label{eq:complex-params}
\lambda_{j} = |\lambda_{j}| \, e^{i\, \theta_{j}} \quad (j = 1,2,3), \quad
\quad
\mbox{and}
\quad
\mu^2_{12} = |\mu^2_{12}| \, e^{i\, \theta_{12}}\,.
\ee
It is important to note that $\lambda_1$, along with other dark-sector parameters $\lambda_{11},\lambda_{22},\lambda_{12}, \lambda'_{12}$, governs only self-interactions among inert scalars and does not affect the tree-level DM or collider phenomenology of the model. These parameters are constrained solely by perturbative unitarity and boundedness-from-below conditions on the scalar potential. As they do not enter our tree-level analysis, their values are fixed to $0.1$.\footnote{The subtleties of the loop-level analysis will be discussed in Sec.~\ref{sec:EWPT}.}

The phenomenologically relevant parameters are $\mu_3^2$ and $\lambda_{33}$, which are fixed by the Higgs mass, and $\mu^2_{1},\,\mu^2_{2},\,\mu^2_{12}, \,\lambda_{31},\, \lambda_{23},\, \lambda'_{31},\,\lambda'_{23},\,\lambda_{2},\,\lambda_{3}$, which determine the inert scalar masses and their couplings. While the latter nine parameters are a priori independent, we simplify the analysis by adopting the \textit{dark democracy} limit~\cite{Keus:2014jha,Keus:2015xya,Cordero-Cid:2016krd,Cordero-Cid:2018man}, in which
\begin{equation}
\label{eq:dark-democracy}
\mu^2_1 = \mu^2_2, \quad \lambda_3 = \lambda_2, \quad \lambda_{31} = \lambda_{23}, \quad \lambda'_{31} = \lambda'_{23}\, .
\end{equation}
The model remains explicitly CP violating in this limit, provided
\begin{equation}
(\lambda_{22} - \lambda_{11}) \left[ \lambda_1 ({\mu^2_{12}}^*)^2 - \lambda^*_1 (\mu^2_{12})^2 \right] \neq 0,
\end{equation}
as shown in~\cite{Haber:2006ue,Haber:2015pua}. This condition ensures that at least one CP-odd invariant is non-zero.\footnote{For a 3HDM to violate CP explicitly, at least one CP-odd invariant must be non-vanishing. The condition above is sufficient but not necessary, as other non-zero invariants can also lead to CP violation.} 
By imposing the dark democracy limit, the only two parameters that remain complex are $\mu^2_{12}$ and $\lambda_2$. Note, however, that one can redefine the doublets as 
\be 
\left\{\begin{array}{c}
\phi_1 \to \phi_1 e^{i \theta_{12}/2} ~\\[2mm]
\phi_2 \to \phi_2 e^{-i \theta_{12}/2}\\[2mm]
\phi_3 \to \phi_3 ~~~~~~~~~
\end{array}
\right.
~~
\Longrightarrow 
~~
\left\{\begin{array}{c}
|\mu^2_{12}|\, e^{i \theta_{12}} \to |\mu^2_{12}|~~~~~~~ \\[2mm]
|\lambda_2|\, e^{i \theta_2}  \to |\lambda_2|\, e^{i (\theta_2+\theta_{12})}
\end{array}
\right.
\ee
to nominally remove the phase of the $\mu^2_{12}$ parameter. As a result, the only relevant CP-violating parameter in the dark democracy limit is the \textit{shifted} phase of the $\lambda_2$ coupling, $\theta_2 + \theta_{12}$, which we denote by $\theta_{\rm CPV}$ throughout the paper.

\subsection{The mass spectrum}
\label{sec:minimisation}
The vacuum configuration $(0, 0, \VSqhalf{v})$ minimises the potential when $v^2 = \mu_3^2 / \lambda_{33}$. The active doublet $\phi_3$ plays the role of the SM Higgs doublet, yielding the massless Goldstone bosons $G^0$ and $G^\pm$, and the SM-like Higgs boson $h$ with
\begin{equation}
m_h^2 = 2\mu_3^2 = 2\lambda_{33} v^2 = (125~\text{GeV})^2.
\end{equation}

\paragraph{The charged inert states:}

The charged mass-squared matrix in the $(H^\pm_1, H^\pm_2)$ basis is given by
\begin{equation}
\mathcal{M}^2_C = 
\begin{pmatrix}
- \mu_2^2 + \frac{1}{2} \lambda_{23} v^2 & -\mu_{12}^2 \\
-\mu_{12}^2 & - \mu_2^2 + \frac{1}{2} \lambda_{23} v^2 
\end{pmatrix}.
\end{equation}
The physical charged eigenstates are
\begin{equation}
S^\pm_1 = \frac{H^\pm_1 + H^\pm_2}{\sqrt{2}}, \qquad
S^\pm_2 = \frac{H^\pm_1 - H^\pm_2}{\sqrt{2}}.
\end{equation}
The corresponding masses are
\begin{equation}
m^2_{S^\pm_1} = -\mu_2^2 - \mu_{12}^2 + \frac{1}{2} \lambda_{23} v^2, \qquad
m^2_{S^\pm_2} = -\mu_2^2 + \mu_{12}^2 + \frac{1}{2} \lambda_{23} v^2,
\end{equation}
where we take $\mu_{12}^2 > 0$, such that $m_{S^\pm_1} < m_{S^\pm_2}$.

\paragraph{The neutral inert states:}

The neutral mass-squared matrix in the $(H_1, H_2, A_1, A_2)$ basis is
\begin{equation}
\mathcal{M}^2_N = \frac{1}{4} \begin{pmatrix}
\Lambda^+_c & -2\mu^2_{12} & -\Lambda_s & 0 \\[2mm]
-2\mu^2_{12} & \Lambda^+_c & 0 & \Lambda_s \\[2mm]
-\Lambda_s & 0 & \Lambda^-_c & -2\mu^2_{12} \\[2mm]
0 & \Lambda_s & -2\mu^2_{12} & \Lambda^-_c
\end{pmatrix},
\label{eq:neutral-mass-squared}
\end{equation}
where
\begin{equation}
\Lambda_s = 2|\lambda_2|  v^2 \sin \theta_{\rm CPV} , \qquad
\Lambda^\pm_c = -2\mu^2_2 + v^2(\lambda_{23} + \lambda'_{23} \pm 2|\lambda_2| \cos \theta_{\rm CPV}) \,.
\end{equation}

The physical CP-mixed neutral eigenstates $S_{1,2,3,4}$ are given by
\begin{align}
S_1 &= \frac{\alpha H_1 - A_1 + \alpha H_2 + A_2}{\sqrt{2(\alpha^2 + 1)}}, &
S_2 &= \frac{H_1 + \alpha A_1 + H_2 - \alpha A_2}{\sqrt{2(\alpha^2 + 1)}}, \\
S_3 &= \frac{\beta H_1 + A_1 - \beta H_2 + A_2}{\sqrt{2(\beta^2 + 1)}}, &
S_4 &= \frac{-H_1 + \beta A_1 + H_2 + \beta A_2}{\sqrt{2(\beta^2 + 1)}},
\end{align}
with mixing parameters
\begin{equation}
\label{eq:alpha-beta}
\alpha = \frac{-\mu^2_{12} + v^2 |\lambda_2| \cos\theta_{\rm CPV} - \Lambda^-}{v^2 |\lambda_2| \sin\theta_{\rm CPV}}, \qquad
\beta = \frac{-\mu^2_{12} - v^2 |\lambda_2| \cos\theta_{\rm CPV} + \Lambda^+}{v^2 |\lambda_2| \sin\theta_{\rm CPV}},
\end{equation}
and
\begin{equation}
\label{eq:lambdas}
\Lambda^\mp = \sqrt{(\mu^2_{12})^2 + v^4 |\lambda_2|^2 \mp 2 v^2 \mu^2_{12} |\lambda_2| \cos\theta_{\rm CPV}}.
\end{equation}

The corresponding masses are
\begin{align}
\label{masses-Ss}
m^2_{S_{1,2}} &= -\mu^2_2 + \frac{v^2}{2}(\lambda'_{23} + \lambda_{23}) \mp \Lambda^-, \\
m^2_{S_{3,4}} &= -\mu^2_2 + \frac{v^2}{2}(\lambda'_{23} + \lambda_{23}) \mp \Lambda^+.
\end{align}
We identify $S_1$ as the lightest inert scalar and the DM candidate. Throughout the paper, we use the notations $S_1$ and DM interchangeably.

\subsection{The span of $\theta_{\rm CPV}$}
\begin{itemize}
\item 
To reproduce the results of~\cite{Keus:2014jha,Keus:2015xya} in which $\lambda_2 < 0$ and $S_1$ is the DM candidate, we require\footnote{For $\lambda_2 > 0$, the same statements hold upon relabelling $S_1 \leftrightarrow S_3$ and $S_2 \leftrightarrow S_4$.}
\begin{equation}
\frac{\pi}{2} < \theta_{\rm CPV} < \frac{3\pi}{2} \qquad \Rightarrow \qquad m_{S_1} < m_{S_2},\, m_{S_3},\, m_{S_4}.
\end{equation}
In contrast, for $\theta_{\rm CPV}$ in the first or fourth quadrants, $S_3$ becomes the lightest neutral inert particle:
\begin{equation}
\left.
\begin{array}{c}
0 < \theta_{\rm CPV} < \frac{\pi}{2} \\[1mm]
\frac{3\pi}{2} < \theta_{\rm CPV} < 2\pi
\end{array}
\right\}
\quad \Rightarrow \quad
m_{S_3} < m_{S_1},\, m_{S_2},\, m_{S_4}.
\end{equation}

\item 
At $\theta_{\rm CPV} = \frac{\pi}{2}, \frac{3\pi}{2}$ points where $\Lambda^+ = \Lambda^-$, a mass degeneracy occurs:
\begin{equation}
\theta_{\rm CPV} = \frac{\pi}{2}, \frac{3\pi}{2}
\quad \Rightarrow \quad
\left\{
\begin{array}{l}
m_{S_1} = m_{S_3}, \\[1mm]
m_{S_2} = m_{S_4}.
\end{array}
\right.
\label{eq:s1s3-degeneracy}
\end{equation}

\item 
The model reduces to the CP-conserving limit for $\theta_{\rm CPV} = 0, \pi$, where $S_{1,3}$ and $S_{2,4}$ become CP eigenstates:
\begin{equation}
\theta_{\rm CPV} = 0, \pi \quad \Rightarrow \quad
\text{CP-conserving limit}: \;
\left\{
\begin{array}{l}
S_{1,3} = \frac{H_1 \pm H_2}{\sqrt{2}}, \\[1mm]
S_{2,4} = \frac{A_1 \pm A_2}{\sqrt{2}}.
\end{array}
\right.
\label{eq:CPC-limit}
\end{equation}

\end{itemize}

\subsection{Input parameters of the model}
\label{sec:input}

We choose the following set of physical observables as independent input parameters:
\begin{equation}
m_{S_1}, \quad m_{S_2}, \quad m_{S^\pm_1}, \quad m_{S^\pm_2}, \quad \theta_{\rm CPV}, \quad g_{h\text{DM}},
\label{eq:input-params}
\end{equation}
where $g_{h\text{DM}} \equiv g_{S_1 S_1 h}$ is the Higgs-DM coupling. The relevant interaction terms in the Lagrangian are
\begin{equation}
\mathcal{L} \supset g_{Z S_i S_j}\,  Z_\mu (S_i \partial^\mu S_j - S_j \partial^\mu S_i)
+ \frac{v}{2} \, g_{S_i S_i h} \, h S_i^2
+ v \, g_{S_i S_j h} \, h S_i S_j
+ v \, g_{S^\pm_i S^\mp_j h} \, h S^\pm_i S^\mp_j \,.
\label{eq:ghSS}
\end{equation}
The parameters of the scalar potential can be expressed in terms of the input set in Eq.~\eqref{eq:input-params} as:
\begin{align}
\mu^2_{12} &= \frac{1}{2}(m^2_{S^\pm_2} - m^2_{S^\pm_1}), \\
\lambda_{23} &= \frac{1}{v^2} \left(2\mu_2^2 + m^2_{S^\pm_2} + m^2_{S^\pm_1} \right), \nonumber \\
\lambda'_{23} &= \frac{1}{v^2} \left( m^2_{S_2} + m^2_{S_1} - m^2_{S^\pm_2} - m^2_{S^\pm_1} \right), \nonumber \\
|\lambda_2| &= \frac{1}{v^2} \left[
\mu^2_{12} \cos\theta_{\rm CPV}
+ \frac{1}{4} \sqrt{
(2 \mu^2_{12} \cos\theta_{\rm CPV})^2
+ (m^2_{S_2} - m^2_{S_1})^2
- (m^2_{S^\pm_2} - m^2_{S^\pm_1})^2
} \right], \nonumber \\
\mu^2_2 &= \frac{v^2}{2} g_{h\text{DM}}
- \frac{v^2}{\alpha^2 + 1}
\left( 2\alpha \sin\theta_{\rm CPV}
+ (\alpha^2 - 1) \cos\theta_{\rm CPV} \right)
|\lambda_2|
- \frac{1}{2}(m^2_{S_2} + m^2_{S_1}). \nonumber
\label{eq:parameters}
\end{align}
We impose the condition that the extracted value of $|\lambda_2|$ must be real and positive, ensuring all input parameters correspond to physically meaningful configurations.
Using the above relations, all other masses and couplings can be derived. For instance, the masses of $S_3$ and $S_4$ are
\begin{equation}
m^2_{S_{3,4}} = m^2_{S_1} + \Lambda^- \mp \Lambda^+, \qquad \text{with} \quad
(\Lambda^+)^2 = (\Lambda^-)^2 + 4 v^2 \mu^2_{12} |\lambda_2| \cos\theta_{\rm CPV}.
\end{equation}
The neutral scalar-gauge couplings are given by
\begin{align}
|g_{Z S_1 S_3}| = |g_{Z S_2 S_4}| &= g^{\text{CPC}}_Z \left( \frac{\alpha + \beta}{\sqrt{\alpha^2 + 1} \sqrt{\beta^2 + 1}} \right), \\
|g_{Z S_1 S_4}| = |g_{Z S_2 S_3}| &= g^{\text{CPC}}_Z \left( \frac{\alpha \beta - 1}{\sqrt{\alpha^2 + 1} \sqrt{\beta^2 + 1}} \right),
\end{align}
where $g^{\text{CPC}}_Z = \frac{e}{2 c_{\theta_W} s_{\theta_W}}$ is the $|g_{ZS_iS_j}|$ value in the CP-conserving limit, with $e$ the electric charge and $c_{\theta_W}$, $s_{\theta_W}$ the cosine and sine of the weak mixing angle. Note that
\begin{equation}
g^2_{Z S_1 S_3} + g^2_{Z S_1 S_4} = \left(g_Z^{\text{CPC}}\right)^2, \qquad
g^2_{Z S_2 S_3} + g^2_{Z S_2 S_4} = \left(g_Z^{\text{CPC}}\right)^2,
\end{equation}
while $g_{Z S_1 S_2} = g_{Z S_3 S_4} = 0$ in the dark democracy limit.

The charged scalar-gauge couplings are similarly obtained:
\begin{align}
|g_{W^\pm S^\mp_1 S_1}| = |g_{W^\pm S^\mp_2 S_2}| &= g^{\text{CPC}}_W \left( \frac{\alpha}{\sqrt{\alpha^2 + 1}} \right), \nonumber \\
|g_{W^\pm S^\mp_1 S_2}| = |g_{W^\pm S^\mp_2 S_1}| &= g^{\text{CPC}}_W \left( \frac{1}{\sqrt{\alpha^2 + 1}} \right), \nonumber \\
|g_{W^\pm S^\mp_1 S_3}| = |g_{W^\pm S^\mp_2 S_4}| &= g^{\text{CPC}}_W \left( \frac{1}{\sqrt{\beta^2 + 1}} \right), \nonumber \\
|g_{W^\pm S^\mp_1 S_4}| = |g_{W^\pm S^\mp_2 S_3}| &= g^{\text{CPC}}_W \left( \frac{\beta}{\sqrt{\beta^2 + 1}} \right),
\end{align}
where $g^{\text{CPC}}_W = \frac{e}{s_{\theta_W}}$ is the $|g_{W^\pm S_{i}^\mp S_j}| $ value in the CP-conserving limit. 
It is important to note that, unlike the CP-conserving case, the gauge-scalar interactions now depend on the mixing parameters $\alpha$ and $\beta$, defined in Eq.~\eqref{eq:alpha-beta}, and therefore on the scalar masses in the CP-violating scenario.

\subsubsection{Renormalization and input parameters}

In this work, thermal corrections to masses, couplings, and the effective potential $\Veff$ are computed at parametric order $g^4$ or $\lambda^2$. In contrast, the $T = 0$ input parameters are only accurate to $\mathcal{O}(g^2)$, as we do not include loop corrections to pole masses or the full matching of renormalised action parameters to physical observables at $\mathcal{O}(g^4)$. This discrepancy should be borne in mind when interpreting results.

Our focus here is primarily on the thermal dynamics of the model, and not on the precision matching of $T = 0$ quantities. A similar treatment, with analogous justifications, was adopted in the context of singlet-extended models in~\cite{Niemi:2021qvp}, where more details can be found.

\section{Constraints on the parameter space}
\label{sec:constraints}

The parameter space of the model is constrained by theoretical, observational and experimental bounds which are satisfied in all our benchmark scenarios to follow:

\begin{enumerate}
\item 
For the potential to have a stable vacuum (i.e. for the potential to be bounded from below), the following conditions are required~\cite{Grzadkowski:2009bt}:
\bea
&& \lambda_{ii}>0, \quad i =1,2,3, \label{positivity1} \\
&& \lambda_x > - 2 \sqrt{\lambda_{11} \lambda_{22}}, \quad \lambda_y > - 2 \sqrt{\lambda_{11} \lambda_{33}}, \quad \lambda_z > - 2 \sqrt{\lambda_{22} \lambda_{33}}, \label{positivity2}\\
&&\left\lbrace  \begin{array}{l} 
\sqrt{\lambda_{33}} \lambda_x + \sqrt{\lambda_{11}} \lambda_z+\sqrt{\lambda_{22}} \lambda_y \geq 0\\
\quad \textrm{or}\\[1mm]
\lambda_{33} \lambda_x^2 + \lambda_{11} \lambda_z^2+\lambda_{22} \lambda_y^2 -\lambda_{11} \lambda_{22} \lambda_{33} - 2 \lambda_x \lambda_y \lambda_z <0.
\end{array}\right.
\label{positivity3}
\eea
where 
\bea
\lambda_x = \lambda_{12}+\textrm{min}(0,\lambda_{12}'-2|\lambda_1|),\\
\lambda_y = \lambda_{31}+\textrm{min}(0,\lambda_{31}'-2|\lambda_3|),\\
\lambda_z = \lambda_{23}+\textrm{min}(0,\lambda_{23}'-2|\lambda_2|).
\eea
As noted in~\cite{Faro:2019vcd}, these conditions are sufficient but
not necessary, as it is possible to construct examples of this model in which the potential is bounded from below, but which violate conditions Eqs.~\eqref{positivity1}-\eqref{positivity3}.

\item
We take all couplings to be $|\lambda_i| \leq\,4\,\pi$ in accordance with perturbative unitarity limits. 

\item 
Parametrised by the electroweak oblique parameters $S,T,U$~\cite{Altarelli:1990zd,Peskin:1990zt,Peskin:1991sw,Maksymyk:1993zm}, inert particles $S_i, S_i^{\pm}$ may introduce important radiative corrections to gauge boson propagators.
We impose a $2\sigma$ agreement with electroweak precision observables at $95 \%$ confidence level~\cite{Baak:2014ora},
\be 
S = 0.05\pm 0.11, \quad T = 0.09\pm 0.13, \quad U = 0.01\pm 0.11.
\ee
Similar to the 2HDM, this condition requires each charged state to be close in mass with a neutral state, in the dark sector~\cite{Dolle:2009fn,Grimus:2007if}.

\item 
The contribution of the inert scalars to the total decay width of the electroweak gauge bosons  constrains the masses of the inert scalars to be~\cite{Agashe:2014kda}
\be 
\label{eq:gwgz}
m_{S_i}+m_{S_{1,2}^\pm}\,\geq\,m_{W^\pm}, \quad
\,m_{S_i}+m_{S_j}\,\geq\,m_Z, \quad
\,2\,m_{S_{1,2}^\pm}\,\geq\,m_Z, \quad
i,j=1,2,3,4.
\ee

\item  
Non-observation of charged scalars puts a model-independent lower bound on their mass~\cite{Lundstrom:2008ai,Cao:2007rm,Pierce:2007ut} and an upper bound on their lifetime~\cite{Heisig:2018kfq} to be
\be 
m_{S_{1,2}^\pm}\,\geq\,70\,\GeV, \qquad
\tau_{S_{1,2}^\pm}\,\leq\,10^{-7} \, s \; \Rightarrow \;
\Gamma^\text{tot}_{S_{1,2}^\pm}\,\geq\,6.58\,\times\,10^{-18}\,\GeV,
\ee
to guarantee their decay within the detector. 
In all our benchmark scenarios, the mass of both charged scalars is above 95 GeV and their decay width, primarily to $S^{\pm}_i \to S_j W^\pm$, is of the order of $10^{-1}$ GeV, which is well within limits.

\item
Any model introducing new decay channels for the SM-Higgs boson is constrained by an upper limit on the Higgs total decay width, $\Gamma^h_\text{tot}\,\leq\,9$ MeV~\cite{CMS:2018bwq}, and Higgs signal strengths~\cite{Khachatryan:2016vau,Aaboud:2018xdt,Sirunyan:2018ouh}. 
In our model, the SM-like Higgs could decay to a pair of inert scalars, provided $m_{S_i}+m_{S_j} < m_h$ and $S_{i,j}$ are long-lived enough ($\tau\,\geq\,10^{-7}$ s). 
As a result, $S_{i,j}$  will not decay inside the detector and therefore contribute to the Higgs invisible decay, $h \to S_i S_j$, with a branching ratio of
\be
\textrm{BR}(h \to S_i S_j) = \frac{\sum_{i,j} \Gamma(h\to S_iS_j)}{\Gamma_h^{\rm SM}+\sum_{i,j} \Gamma(h \to S_iS_j)}, 
\label{inv_all}
\ee
with
\be
\Gamma(h\to S_i S_j)=\frac{g_{h S_i S_j}^{2}v^2}{32\pi m_{h}^3}
\biggl( \left(m_h^2-(m_{S_i}+m_{S_j})^2 \right) \left(m_h^2-(m_{S_i}-m_{S_j})^2 \right)\biggr)^{1/2},
\ee
which sets strong limits on the Higgs-inert couplings. 
Moreover, the partial decay $\Gamma(h\to \gamma\gamma)$ receives contributions from the inert charged scalars. 
The combined ATLAS and CMS Run I results for Higgs to $\gamma\gamma$ signal strength require $\mu_{\gamma \gamma} = 1.14^{+0.38}_{-0.36}$~\cite{Khachatryan:2016vau}. 
In Run II, ATLAS reports $\mu_{\gamma \gamma} = 0.99^{+0.14}_{-0.14}$~\cite{Aaboud:2018xdt}, and CMS reports $\mu_{\gamma \gamma} = 1.18^{+0.17}_{-0.14}$ \cite{Sirunyan:2018ouh} with both of which we are in  $2\sigma$  agreement.

\item
Reinterpretation of LEP 2 and LHC Run I searches for Supersymmetric (SUSY) particles (mainly sneutrinos and sleptons) for the IDM ~\cite{Lundstrom:2008ai,Belanger:2015kga} excludes the region of parameter space where the following conditions are simultaneously satisfied ($i=2, 3, 4 $):
\be 
\label{eq:leprec}
m_{S_1}\,\leq\,80\,\GeV,\,\, ~~
m_{S_i}\,\leq\,100\,\GeV,\,~~
\Delta m {(S_1,S_i)}\,\geq\,8\,\GeV.
\ee
We take these limits into account for our DM candidate paired with any other neutral scalar.
We also check the validity of our benchmark scenarios against LHC searches for new particles in accordance with the analysis for the IDM~\cite{Kalinowski:2018ylg}.

\item 
DM relic density measurements from the Planck experiment~\cite{Ade:2015xua},
\be
\label{eq:planck}
\Omega_{\mathrm{DM}}\,h^2\,=\,0.1197\,\pm\,0.0022,
\ee
require the relic abundance of the DM candidate to lie within these bounds if it constitutes 100\% of DM  in the universe.

A DM candidate with $\Omega_{\rm DM} h^2 $ smaller than the observed value is allowed; however, an additional DM candidate is needed to complement the missing relic density. Regions of the parameter space corresponding to values of $\Omega_{\rm DM}h^2$ larger than the Planck upper limit are excluded.

We impose a $3\sigma$ agreement with the observation on the relic abundance of our DM candidate, $S_1$.

\item 
The latest XENON1T results for DM direct detection experiments~\cite{Aprile:2018dbl} and FermiLAT results for indirect detection searches~\cite{Fermi-LAT:2016uux} do not constrain the model any further. Having set the Higgs portal couplings to zero in our benchmark scenarios, the largest direct detection cross section is $\sigma_{DM-N} \approx 10^{-14}$ pb and the largest indirect detection cross section is $\langle v\sigma \rangle \approx 10^{-32}$ cm$^3/$s, both of which are well below the limits~\cite{Billard:2013qya}.

\end{enumerate}

\subsection{DM abundance and the selection of benchmarks}
\label{selection}

The relic abundance of the DM candidate, $S_1$, after freeze-out is given by the solution of the Boltzmann equation,
\be 
\frac{d n_{S_1}}{dt} = 
- 3\, H \,n_{S_1} - \langle \sigma_{eff}\, v \rangle \, \left[ (n_{S_1})^2 - (n^{eq}_{S_1})^{2} \right],
\ee
where $n_{S_1}$ is the number density of the $S_1$ particle, $H$ is the Hubble parameter, and $n^{eq}_{S_1}$ is the number density of $S_1$ at equilibrium. 
The thermally averaged effective (co)annihilation cross section, $\langle \sigma_{eff}\, v \rangle$, receives contributions from all relevant annihilation processes of any $S_i S_j$ pair into SM particles, so that
\be 
\langle \sigma_{eff} v \rangle = 
\sum_{i,j} \langle \sigma_{ij}\, v_{ij} \rangle \,\frac{n^{eq}_{S_i}}{n^{eq}_{S_1}} \, \frac{n^{eq}_{S_j}}{n^{eq}_{S_1}},
\qquad
\mbox{where}
\qquad
\frac{n^{eq}_{S_i}}{n^{eq}_{S_1}} \sim \exp({-\frac{m_{S_i} - m_{S_1}}{T}}).
\ee
However, only processes with the ${S_i}-{S_1}$ mass splitting comparable to the thermal bath temperature $T$ provide a sizeable contribution.

A common feature of non-minimal Higgs DM models is that in a large region of the parameter space the most important process for DM annihilation is through the
$S_1 S_1 \to h_{\rm SM} \to f \bar f$
channel whose efficiency depends on both the DM mass and the Higgs-DM coupling. 
In the region where $m_{\rm DM} < m_h/2$, generally one requires a large Higgs-DM coupling in order to produce relic density in agreement with Eq.~\eqref{eq:planck}.
However, such large Higgs-DM coupling leads to large direct detection and indirect detection cross sections and significant deviations from SM-Higgs coupling measurements, which are ruled out by experimental and observational data. 
On the other hand, a small Higgs-DM coupling fails to annihilate the DM candidate effectively and leads to the over-closure of the universe.
This is where co-annihilation processes play an important role as they can contribute to changes in the DM relic density.
 
In models with extended dark sectors, in addition to the standard Higgs mediated annihilation channels of DM, there exists the possibility of co-annihilation with heavier states, provided they are close in mass~\cite{Cordero-Cid:2016krd,Cordero:2017owj,Cordero-Cid:2018man,Keus:2014jha,Keus:2015xya}. The relevance of this effect depends not only on the DM mass and the mass splittings but also on the strength of the standard DM annihilation channel. 

It is worth pointing out that in the IDM, where by construction CP violation is not allowed, the only co-annihilation process is through the $Z$-mediated $H\,A \to Z \to f \bar f$ channel whose sub-dominant effect fails to revive the model in the low mass region. 
Extending the inert sector,
as shown in~\cite{Cordero:2017owj,Keus:2014jha,Keus:2015xya} in the CP-conserving limit, opens up several co-annihilation channels, both Higgs-mediated $H_1\,H_2 \to h \to f \bar f$ and $Z$-mediated $H_1\,A_{1,2} \to Z \to f \bar f$. However, their collective contribution to DM co-annihilation is not sufficient and one still needs a non-zero Higgs-DM coupling to satisfy relic density bounds.
Introducing CP violation in the extended dark sector~\cite{Cordero-Cid:2016krd,Cordero-Cid:2018man,Fuyuto:2019vfe} opens up many co-annihilation channels through the Higgs and $Z$ bosons, $S_i\,S_j \to h/Z \to f \bar f$, which can significantly affect the DM phenomenology.
In fact, the $Z$-mediated co-annihilations can be strong enough to relieve the model of the need for any Higgs-mediated (co)annihilation processes. 

To show the effect of $Z$ portal CP violation on the abundance of DM, we set the Higgs-DM coupling to zero, $g_{h\mathrm{DM}}=0$, thereby removing the main DM annihilation process, $S_1 S_1 \to h \to f \bar f$. All other $S_iS_j h$ vertex coefficients are also reduced to a point where their resulting co-annihilation processes have negligible contributions to the DM relic density. So, the only communication between the dark sector and the visible sector is through the gauge bosons $W^\pm$ and $Z$ .

In the region of the parameter space where Higgs portal interactions are negligible ($g_{h\mathrm{DM}} \approx 10^{-4}$), the total DM annihilation cross section receives contributions from the following:
\begin{itemize}
\item 
\textbf{DM annihilation processes:}
\be 
S_1 S_1 \to V V , \qquad
S_1 S_1 \to V V^* \to V f f', \qquad 
S_1 S_1 \to V^* V^* \to f f' f f' \,,
\label{annihilation-1}
\ee
where $V$ is any of the SM gauge bosons. In the $m_{DM} < m_{W^\pm}$ region, the processes with off-shell gauge bosons dominate over the ones with on-shell gauge bosons.

\item 
\textbf{DM co-annihilation processes:}
\be 
S_1 S_{2,3,4} \to Z^* \to f \bar f, \qquad S_1 S^\pm_{1,2} \to W^{\pm *} \to f f' \, ,
\label{annihilation-2}
\ee
where the co-annihilating dark scalars are up to 20\% heavier than the DM candidate.

\item 
\textbf{(co)annihilation of other dark states:}
\be  
S_i S_i \to V V , \quad
S_i S_i \to V V^* \to V f f', \quad 
S_i S_i \to V^* V^* \to f f' f f', \quad 
S_i S_j \to V^* \to f f', 
\label{annihilation-3}
\ee
where $S_i\neq S_j$ are any of the dark scalars $ S_{2,3,4}\,, S^\pm_{1,2}$ which are all close in mass.
\end{itemize}
Taking all such processes into account, we define our benchmark scenarios in Sec.~\ref{sec:EWPT} with distinct DM phenomenology.
It is convenient to introduce the mass splittings between the DM candidate and other inert scalars as
\be
\delta_{12} = m_{S_2} - m_{S_1}, \qquad \delta_{c} = m_{S_2^\pm} - m_{S_1^\pm} , \qquad \delta_{1c} = m_{S_1^\pm} - m_{S_1}\, .
\ee

\section{Thermal corrections}
\label{sec:thermal}

To study the phase structure of the model at finite-temperature we construct the effective potential, $\Veff$, using the \textit{dimensional reduction} formalism briefly outlined as follows. Dimensional reduction starts with the imaginary time formalism of finite-temperature field theory and where the (imaginary) time integral in the action spans from 0 to ${1}/{T}$ in order to marry the partition functions of quantum field theory (QFT) and thermal and statistical physics (TSP),
\be 
\label{eq:Partiction-function}
Z_{QFT} =  \mathrm{Tr}\left[e^{-it \hat{H}} \right] \qquad \mbox{and} \qquad
Z_{TSP} = \mathrm{Tr}\left[e^{-\hat{H}/T}\right]  \,.
\ee
This compactification leads to Matsubara modes $\omega_n$~\cite{Matsubara:1955ws} which are discrete modes that can be easily seen in the propagator:
\be 
\sum_{n=-\infty}^{\infty} \int \text{d}^3 p \left( \frac{1}{ {\vec{p}}^{\,2} + m^2 + w^2_n} \right), \quad \begin{cases} 
\omega_n = 2 n \pi T & \text{for bosons} \\
\omega_n = (2 n +1) \pi T & \text{for fermions} 
\end{cases}
\ee 
In the high temperature limit, $\pi T \gg m$, where $m$ denotes the particle mass of the particle the non-zero Matsubara modes, otherwise known as \textit{hard modes}, are much heavier than the zero modes, i.e. \textit{soft modes}, and so can be integrated out. This leaves just the soft modes of the bosonic sector to govern the infrared (IR) behaviour of the effective theory.
These soft modes have no (imaginary) time dependence, leading to a trivial time integration which reduces the theory from a four-dimensional (4D) to a three-dimensional (3D) effective field theory (EFT),
\be 
\label{eq:DR-action}
S = \int_0^{\frac{1}{T}} \text{d} \tau \int \text{d} x^3 \mathcal{L}^{\mathrm{4D}}_E(\tau, \vec{x}) ~~ \xrightarrow{\text{DR}} ~~ \frac{1}{T} \int \text{d} x^3 \mathcal{L}^{\mathrm{3D}}_E(\vec{x}) \,.
\ee 
As a result, the soft modes can be used to describe equilibrium dynamics and are well suited to be used for finding the critical temperature of the phase transition.\footnote{Once the bubbles of the new phase start nucleating and releasing energy into the thermal bath, local equilibrium can be lost. Calling into question the reliability of this approach (see~\cite{Gould:2021ccf} for further details).} 

The light fields that are left in the effective theory, receive large IR contributions from so the so-called \textit{daisy} or \textit{ring} diagrams which are a class of diagrams all of the same order, $\mathcal{O}(g^3 T^4)$ or equivalently $\mathcal{O}(\lambda^{3/2} T^4$). These diagrams require resummation~\cite{Carrington:1991hz, Arnold:1992rz} which is automatically handled in the dimensional reduction treatment.

These IR contributions are particularly important for EWPTs which typically involve light scalar excitations near the critical temperature (the temperature at which the symmetry preserving and broken minima are degenerate).
A more severe issue arises in the gauge sector; in the symmetric phase, gauge bosons remain perturbatively massless, and their spatial components, corresponding to the zero Matsubara modes, become strongly coupled at the scale $g^2 T$. 
This is the origin of the \textit{Linde problem}~\cite{Linde:1980ts} which states that in non-Abelian gauge theories, while the temporal components of the gauge fields acquire a thermal Debye mass of order \( gT \), the spatial components remain massless to all orders in perturbation theory. As a result, perturbative expansions for thermodynamic quantities such as the free energy suffer from severe infrared divergences. In particular, contributions to the free energy at order \( g^6 T^4 \) and beyond become non-perturbative, rendering standard resummation techniques insufficient. This infrared sensitivity necessitates the use of non-perturbative methods, such as dimensional reduction, for an effective 3D theory followed by lattice simulations.


Despite this breakdown at high orders, the lower-order terms in $\Veff$ can be computed reliably within perturbation theory, provided that IR divergences are under control. The commonly used one-loop effective potential with daisy resummation, as implemented in codes such as \texttt{CosmoTransitions}~\cite{Wainwright:2011kj}, corresponds to a partial $\mathcal{O}(g^3)$ computation(including some $\mathcal{O}(g^4)$ terms from $T = 0$ corrections).\footnote{Note that the power counting described here is schematic, and that for scalars undergoing a phase transition the relevant perturbative expansion can have more a intricate structure~\cite{Ekstedt:2022zro,Gould:2023ovu}.} While qualitatively useful, this approximation is known to substantially overestimate the strength of the EWPT and yield an incorrect value for the critical temperature compared to non-perturbative results~\cite{Kainulainen:2019kyp,Niemi:2020hto,Gould:2022ran,Niemi:2024axp}.
More accurate predictions are obtained at full $\mathcal{O}(g^4)$, which requires two-loop computations and a consistent resummation beyond the daisy level~\cite{Arnold:1992rz,Laine:2017hdk}. 

This can be achieved systematically through the dimensional reduction formalism~\cite{Ginsparg:1980ef,Appelquist:1981vg,Arnold:1992rz,Kajantie:1995dw}. This is done 
by integrating out heavy modes (including non-zero Matsubara frequencies), hence, mapping the 4D theory onto a 3D EFT valid for scales $\lesssim T$. This procedure resums thermal contributions for light fields and organises corrections in powers of $m/T$ and coupling constants. The resulting EFT simplifies both perturbative and non-perturbative treatments of the phase transition. 
In particular, the thermodynamics of the 3D EFT can be studied non-perturbatively on the lattice. This approach has been successfully applied to the SM and its extensions~\cite{Kajantie:1995kf,Kajantie:1996qd,Csikor:1998eu,Laine:1998qk,Kainulainen:2019kyp,Niemi:2020hto,Gould:2022ran,Niemi:2024axp,Niemi:2021qvp}, and has recently been applied to BSM scenarios using automated tools such as the \texttt{DRalgo} package~\cite{Ekstedt:2022bff}. In this work, we employ dimensional reduction to construct the high-$T$ EFT for a 3HDM with a CP-violating dark sector, and compute the two-loop $\Veff$ to determine the nature of the EWPT.

\subsection{High-\texorpdfstring{$T$}{T} effective theory for 3HDMs}


For the 3HDM potential defined in Eq.~\eqref{eq:V0-3HDM}, the high-$T$ EFT has the form
\begin{align}
\label{eq:3D-EFT}
S_{\mathrm{3D}} = \frac{1}{T} \int d^3x \; \Big\{ \frac12 \Tr F_{ij}F_{ij} + \frac14 f_{ij} f_{ij} + \sum_{n=1}^3 (D_i\phi_n)^\dagger (D_i\phi_n) + \bar{V}_\mathrm{3HDM} \Big\},
\end{align}
where $F_{ij}$ and $f_{ij}$ are the $\text{SU}(2)$ and $\text{U}(1)$ field strengths, $D_i$ is the 3D covariant derivative, and $\bar{V}_\mathrm{3HDM}$ has the same form as the 4D potential but with temperature-dependent parameters. Fermions are integrated out during dimensional reduction, so their effects are encoded in the EFT parameters. The explicit $1/T$ prefactor can be absorbed via field rescaling~\cite{Niemi:2021qvp}, but is kept here for clarity.

We assume all physical masses are parametrically small compared to the temperature, including the scalar mixing parameter $\mu_{12}^2$. Without this assumption, diagonalising the $\phi_1$–$\phi_2$ subspace could yield heavy eigenstates with $m \gg T$, invalidating the high-$T$ assumption required for constructing the EFT. We therefore adopt the power counting $\mu_{12}^2 \sim g^2 T^2$ (or alternatively $\mu_{12}^2 \sim \lambda T^2$), allowing the $\mu_{12}^2$ mixing term to be treated as a perturbative two-point interaction in Feynman diagrams. This treatment is consistent with Refs.~\cite{Losada:1996ju,Bodeker:1996pc,Andersen:1998br,Laine:2000rm,Gorda:2018hvi} and is implemented in \texttt{DRalgo}.

Several comments on the structure and truncation of the EFT are in order:
\begin{itemize}
	\item 
    As mentioned before, the symmetry of the potential in Eq.~\eqref{eq:V0-3HDM} allows for operators of the form $(\phi_1^\dagger\phi_2)(\phi_1^\dagger\phi_1)$,
	$(\phi_1^\dagger\phi_2)(\phi_2^\dagger\phi_2)$,
	$(\phi_3^\dagger\phi_1)(\phi_2^\dagger\phi_3)$,
	$(\phi_1^\dagger\phi_2)(\phi_3^\dagger\phi_3)$
	and their conjugates. Since we do not expect these terms to qualitatively change the phenomenology of the model, we have set their coefficients to zero at tree level. In principle, these operators will be generated by thermal loops and should be included in the EFT. However, they can only arise from loops that contain the $\mu_{12}^2 \, \phi^{\dagger}_1 \phi_2$ interaction vertex and are thus suppressed in the high-$T$ approximation. The lowest order at which these operators must be included is $\lambda^2 \mu_{12}^2 \sim \mathcal{O}(\lambda^2 g^2)$ which goes beyond our $g^4$ accuracy goal. We therefore neglect these operators from our EFT. 
	
	\item 
    Mixed kinetic terms such as $\left(D_i\phi_1 \right)^\dagger \left(D_i\phi_2 \right)$ could similarly be  generated by loop corrections arising from momentum dependence of two-point correlators of the doublets in a quadratic $\lambda^2 \mu_{12}^2 \sim \mathcal{O}(\lambda^2 g^2)$ vertex. The simplest contributing diagram is of the order $g^2 \, \mu_{12}^2 \, p^2$, where $p$ is an external momentum at the EFT scale. Since the relevant momentum scale for the EFT is $p \leq gT$, kinetic mixing operators do not contribute at order $g^4$.
	
	\item 
    As with any local EFT, the integration over heavy modes generates operators at all possible field dimensions. However, our constructed EFT in Eq.~\eqref{eq:3D-EFT} is truncated at field dimension four. In this setup, the first corrections to the EFT arise at dimension six and include operators such as $(\phi_3^\dagger\phi_3)^3$. The Wilson coefficients associated with dimension-six operators are formally of order $g^6$ and can again be neglected in our $g^4$ analysis. It is important to note that this counting breaks down if the high-$T$ assumption, $m \ll \pi T$, is not strictly satisfied. Violating this assumption could lead to higher-dimensional operators becoming important if there are large mass hierarchies in the scalar sector. Effects of thermal dimension-six operators in the context of EWPT, have been studied in~\cite{Kajantie:1995dw, Niemi:2018asa, Gould:2019qek, Niemi:2021qvp, Niemi:2024vzw}.
		
	\item 
    In constructing the EFT in Eq.~\eqref{eq:3D-EFT}, we have also integrated out the time component of gauge fields ($A_0, B_0$, and the corresponding $\text{SU}(3)$ field), since in the imaginary time formalism, they acquire thermal Debye masses $\sim gT$ and behave like adjoint-representation scalars. We perform the integration over these fields in a separate step after the actual 4D $\rightarrow$ 3D matching giving a theory at the ultrasoft scale ($g^2T)$. This modifies the EFT parameters somewhat, but the numerical impact of this process is known to be sub-leading compared to the top-quark loop effects~\cite{Kajantie:1995dw,Bodeker:1996pc,Niemi:2021qvp}.
    
\end{itemize}

\subsection{The effective potential}

The two-loop effective potential is computed from the EFT in Eq.~\eqref{eq:3D-EFT}, in the $R_\xi$ Landau gauge. We consider homogeneous background fields in the neutral component of each doublet:
\begin{align}
\label{eq:background_fields}
\phi_i \rightarrow \phi_i + \frac{1}{\sqrt{2}} \begin{pmatrix} 0 \\ \bar{v}_i \end{pmatrix},
\end{align}
where $\bar{v}_i$ are real background fields. 
The effective potential is a function of these fields, $\Veff = \Veff(\bar{v}_1, \bar{v}_2, \bar{v}_3)$, and is calculated from one- and two-loop vacuum diagrams (i.e. no external legs) with propagators and vertices depending on $\bar{v}_i$. For details, see e.g.~\cite{Farakos:1994kx,Niemi:2020hto}.

The three-field parametrization used here is not the most general background-field configuration possible in the 3HDM, which includes nine real degrees of freedom after gauge fixing. In particular, configurations with non-vanishing charged or CP-violating background fields are not considered here. 
Such configurations are not expected to dominate the thermal evolution in the regions we study.
The current paper is intended to be a proof of concept that strong first order EWPT is viable in a 3HDM with a CP-violating dark sector. A comprehensive analysis of the 3HDM phase structure, including non-vanishing charged and CP-violating background fields, will be presented in a forthcoming publication.

The full two-loop $\Veff$ can be obtained, in symbolic form, using \texttt{DRalgo}. A complication here is that the loop calculation needs to be done in field-basis where the mass matrix is diagonal, i.e. there are no quadratic mixing terms.\footnote{\texttt{DRalgo} has an option for computing the $\Veff$ using undiagonalized propagators, i.e. all off-diagonal elements in the mass matrix are treated as perturbations. We have not used this approximation in the present paper.}
In practice, we must give \texttt{DRalgo} rotation matrices that diagonalize both the scalar and gauge sectors, and this diagonalization now depends on the background fields $\bar{v}_i$. While for gauge fields this is straightforward and proceeds much like in the minimal SM, for our scalars the diagonalizing rotation cannot be found analytically. Specifically, the 3HDM has $3 \times 4 = 12$ real scalar fields so the mass matrix is $12 \times 12$. For the background-field configuration defined in Eq.~\eqref{eq:background_fields}, it is possible to permute the fields such that the mass matrix becomes block diagonal, reducing the problem to diagonalization of two $6 \times 6$ symmetric matrices which still needs to be done numerically.
Details of our implementation, including the treatment of background-dependent rotation matrices and the \texttt{DRalgo} model files, are available in an open source code that we constructed to numerically evaluate the effective potential, called \texttt{BLOOP (Beyond one LOOp Phase transition)}\footnote{The code is available visa \hyperlink{https://github.com/BLOOP-JTC/BLOOP}{\texttt{https://github.com/BLOOP-JTC/BLOOP}} with the manual upcoming.}.

\section{EWPT dynamics}
\label{sec:EWPT}

\subsection{Setup}

We now outline the numerical setup used to compute the effective potential and extract the phase structure, closely following the methodology of~\cite{Niemi:2024vzw}. As discussed, reliable predictions for the critical temperature of the EWPT require going beyond the standard one-loop approximation. We therefore compute the thermal effective potential at two-loop level using \texttt{DRalgo v1.2} in combination with \texttt{GroupMath v3}~\cite{Fonseca:2020vke} and \texttt{BLOOP v0.1}.\footnote{Only scalar self-energies are computed at two-loop; cubic and quartic couplings are computed at one-loop.} Matching relations, $\beta$-functions, and tree-level mass matrices are also generated via \texttt{DRalgo}. A fully consistent loop-level mass matrix is left for future work.

Matching from the hard-to-soft scale is performed at $\mu = 4\pi e^{-\gamma} T$, where $\gamma$ is the Euler–Mascheroni constant, and the soft-to-ultrasoft scale is matched at $\mu = T$. We minimise the potential with respect to three real CP-even background fields,
\begin{align}
\phi_i = \doublet{$\begin{scriptsize}$ 0 $\end{scriptsize}$}{\frac{\bar{v}_i+0}{\sqrt{2}}}\,,
\end{align}
restricting to CP- and charge-conserving field configurations. This simplification allows the $12\times12$ scalar mass matrix to be permuted into two $6\times6$ blocks for neutral and charged states, respectively. 


We scan the field space over
\begin{align}
v_1 \in (-60, 60)\GeV^{1/2}, \quad v_{2,3} \in (10^{-4}, 60)\GeV^{1/2},
\end{align}
exploiting gauge and $\mathbb{Z}_2$ symmetry to fix $v_2, v_3 > 0$.

Minimisation is performed using the \texttt{BOBYQA} and \texttt{DIRECT} routines from the \texttt{NLOPT} library~\cite{NLOPt}. 
We treat $\Veff$ as a complex-valued function and minimise its real part, using the imaginary part as a consistency check. All global minima are found to be real, indicating absence of spurious solutions. The global minimum is tracked over a temperature range $T \in [50, 400]\,\GeV$ with step size $\Delta T = 0.1\,\GeV$.

We perform a random scan over one million benchmark points with input parameters sampled uniformly in the following ranges:
\begin{align}
\begin{tabular}{c@{\hskip 2em}c}
$m_{S_1} \in [63, 100]\,\GeV$ & $\delta_{12} \in [5, 100]\,\GeV$ \\
$\delta_{1c} \in [5, 100]\,\GeV$ & $\delta_c \in [5, 100]\,\GeV$ \\
$g_{h\mathrm{DM}} \in [0, 1]$ & $\theta_{\mathrm{CPV}} \in [{\pi}/{2}, {3\pi}/{2}]$
\end{tabular}
\label{eq:ScanRange}
\end{align}

Each benchmark is required to satisfy the theoretical and experimental constraints listed in Sec.~\ref{sec:constraints} in addition to ensuring that the global minimum at $T=0$ is at $(0,0,246\,\GeV)$. It is important to note that for most of our benchmark points, specially for points with a relatively large $g_{h\mathrm{DM}}$, the relic density of the DM candidate is only a fraction of the observed relic abundance, which subsequently relaxes the (in)direct detection bounds considerably.

These constraints are conservative; e.g.\ we impose absolute vacuum stability, whereas metastability would suffice. We also reject points where the correct vacuum is not recovered at $T=50\,\GeV$, to avoid scenarios where the transition temperature lies below our scan window.

\subsection{Numerical results}

\begin{figure}[h!]
\centering
{\includegraphics[width=0.49\linewidth]{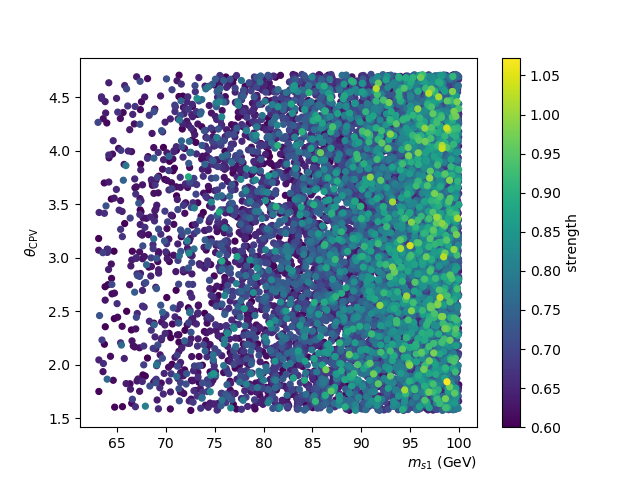}}
{\includegraphics[width=0.49\linewidth]{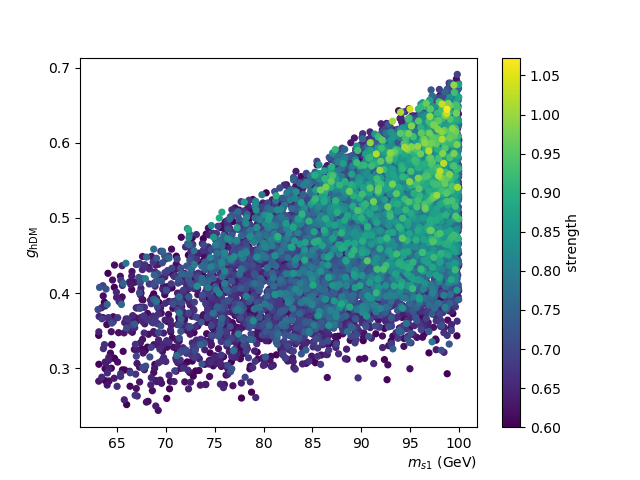}}
{\includegraphics[width=0.49\linewidth]{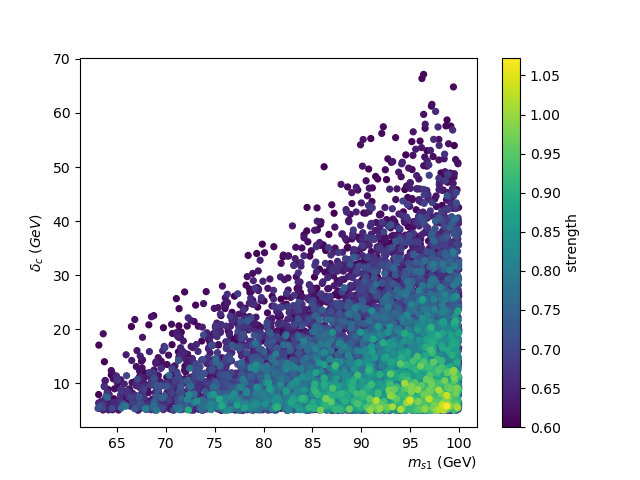}}
{\includegraphics[width=0.49\linewidth]{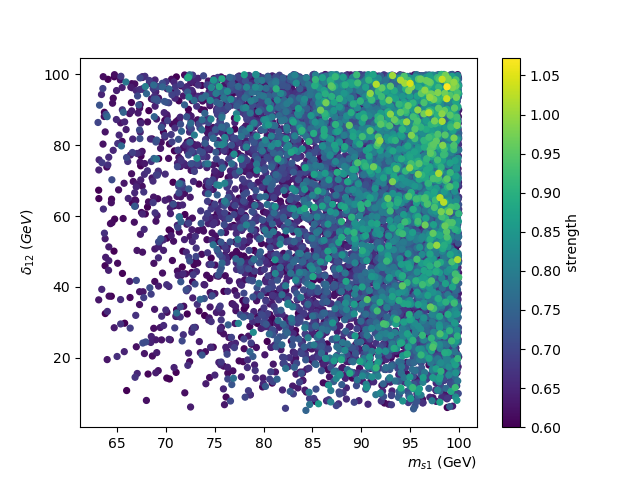}}
{\includegraphics[width=0.49\linewidth]{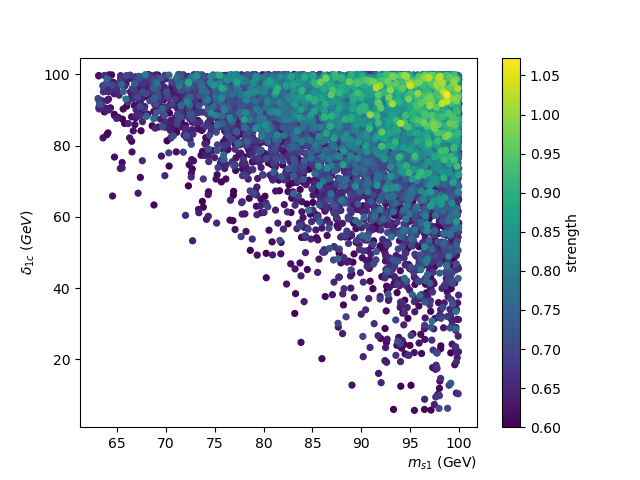}} 
\caption{Heat maps of the phase transition strength at one-loop as a function of scan parameters. The step size is $0.1\,\GeV$.}
\label{fig:ScanFigures-1loop}
\end{figure}

\begin{figure}[h!]
\centering
{\includegraphics[width=0.49\linewidth]{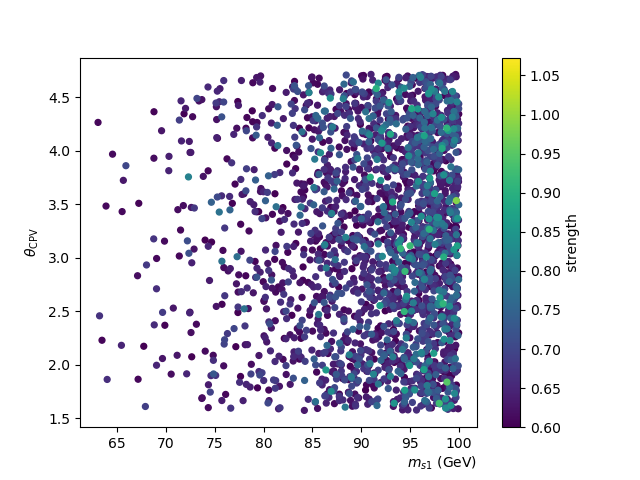}}
{\includegraphics[width=0.49\linewidth]{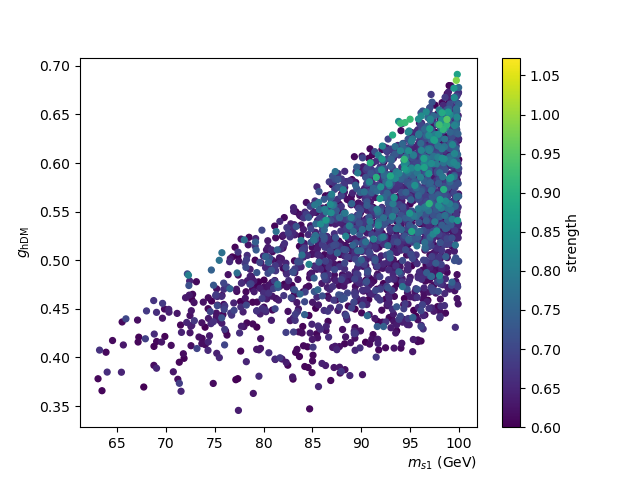}}
{\includegraphics[width=0.49\linewidth]{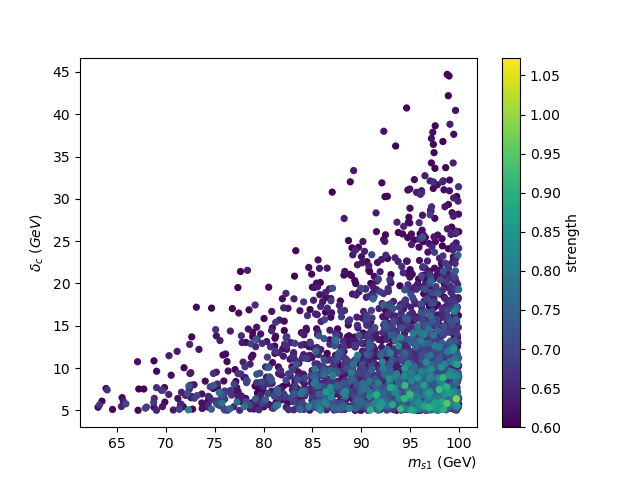}}
{\includegraphics[width=0.49\linewidth]{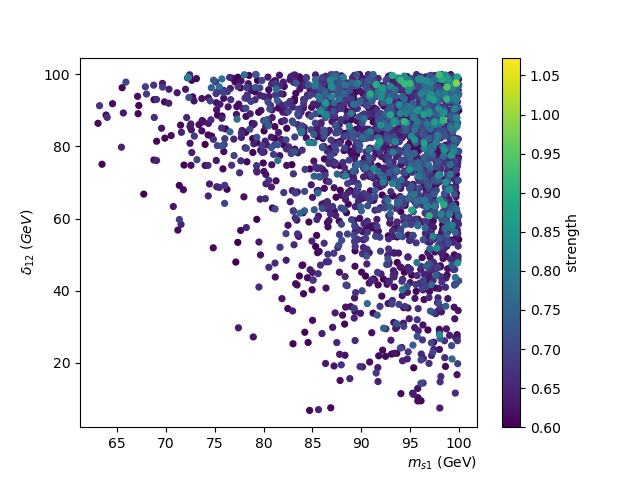}}
{\includegraphics[width=0.49\linewidth]{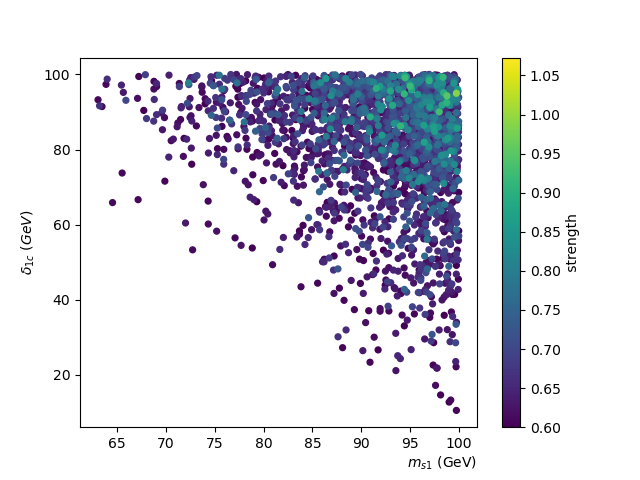}}
\caption{Heat maps of the phase transition strength at two-loop as a function of scan parameters. The step size is $0.1\,\GeV$.}
\label{fig:ScanFigures-2loop}
\end{figure}

We characterise the strength of the phase transition by the dimensionless jump in the Higgs VEV at $T_c$:
\begin{align}
\frac{\Delta v}{\sqrt{T_c}} = \frac{\Delta\sqrt{v_1^2 + v_2^2 + v_3^2}}{\sqrt{T_c}}.
\end{align}
This quantity is used throughout as a proxy for the order parameter. Figs.~\ref{fig:ScanFigures-1loop}-\ref{fig:ScanFigures-2loop} shows the dependence of transition strength on model parameters. White regions correspond to benchmarks with no strong first-order phase transition, either due to insufficient strength or exclusion by constraints. The majority of transitions proceed from the symmetric vacuum to $(0,0,v_3)$; the subset involving two-step transitions is discussed in Sec.~\ref{sec:TwoStep}.

We observe that the strength of the transition increases with $m_{S_1}$ and $g_{h\mathrm{DM}}$, and is largely insensitive to $\theta_{\mathrm{CPV}}$. This trend is consistent with other SM-like models: a strong transition typically requires heavy states coupled to the Higgs to enhance thermal corrections~\cite{Ekstedt:2024etx}. At two-loop level, the transition strength is reduced on average by 27\%, and the critical temperature shifts downward by 7.2\%. Out of one million points, 10,423 exhibit a strong first-order phase transition at one-loop, but only 2,335 remain strong at two-loop.

\begin{figure}[h!]
\centering
{\includegraphics[width=0.49\linewidth]{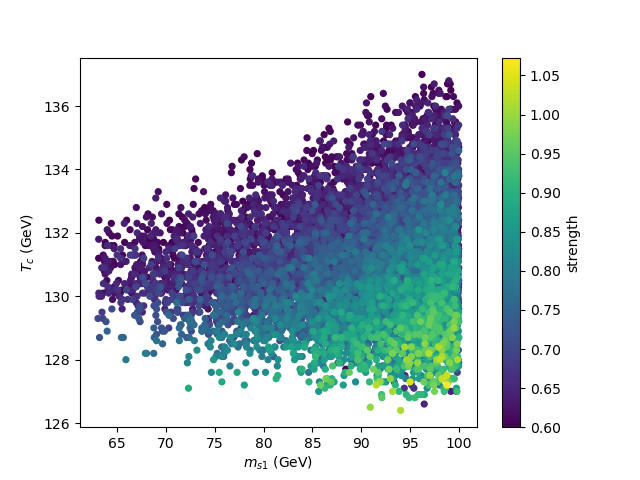}}
{\includegraphics[width=0.49\linewidth]{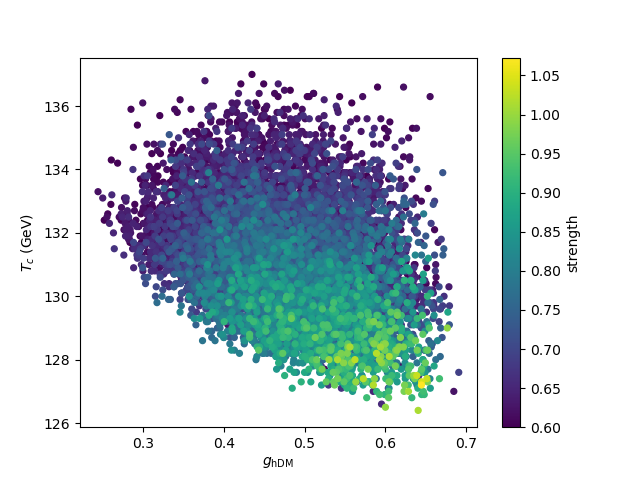}}\\
{\includegraphics[width=0.49\linewidth]{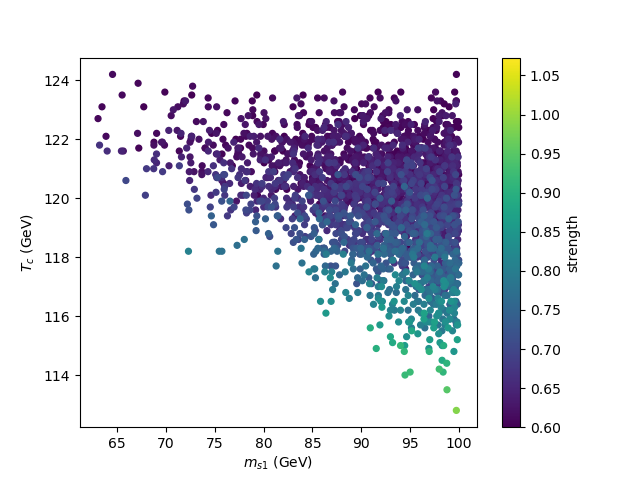}}
{\includegraphics[width=0.49\linewidth]{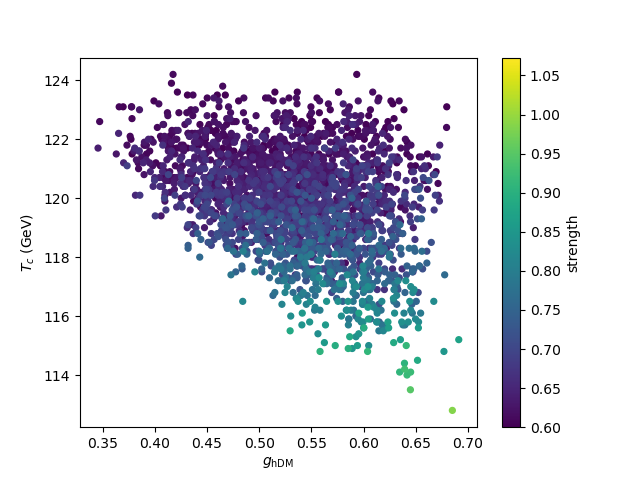}}
\caption{Heat maps of critical temperature $T_c$ as a function of $m_{S_1}$ and $g_{h\mathrm{DM}}$ at one-loop (top) and two-loop (bottom).}
\label{fig:ScanFiguresTc}
\end{figure}

\subsubsection{Towards electroweak baryogenesis}

\begin{figure}[h!]
\centering
{\includegraphics[width=0.49\linewidth]{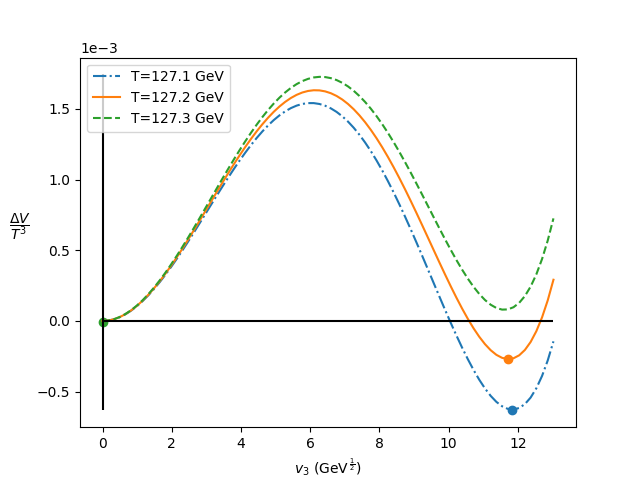}}
{\includegraphics[width=0.49\linewidth]{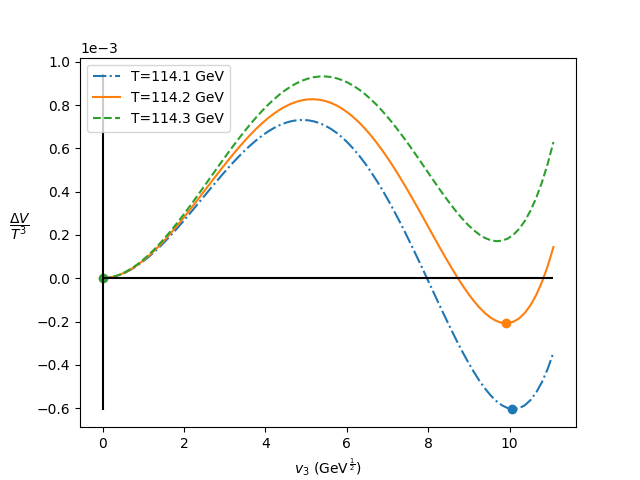}}
\caption{Temperature-normalised effective potential at one-loop (left) and two-loop (right) for the benchmark in Eq.~\eqref{eq:MaximalCPV}. Filled circles indicate the global minimum.}
\label{fig:MaxCPVplot}
\end{figure}

For electroweak baryogenesis, a strong first-order transition must be accompanied by sizeable CP violation, i.e.\ $\theta_{\mathrm{CPV}} \approx \pi/2$. The best candidate from our scan is:
\begin{align}
\theta_{\mathrm{CPV}} &= 1.64, \quad &\delta_{12} &= 99.9\,\GeV, \quad &m_{S_1} &= 98.0\,\GeV, \nonumber\\
g_{h\mathrm{DM}} &= 0.639, \quad &\delta_{1c} &= 90.0\,\GeV, \quad &\delta_c &= 7.47\,\GeV.
\label{eq:MaximalCPV}
\end{align}
This point exhibits an increase in transition strength from 0.78 (one-loop) to 0.93 (two-loop). The corresponding potential profiles are shown in Fig.~\ref{fig:MaxCPVplot}. A dedicated analysis of baryogenesis dynamics in this model will be presented in a forthcoming publication.

\subsubsection{Two-step phase transitions}
\label{sec:TwoStep}

Approximately 20\% of benchmarks exhibit a two-step phase transition at one-loop level. These proceed via an intermediate phase $(v_1, v_2, 0)$, followed by a transition to the electroweak vacuum $(0,0,v_3)$. However, since the first step is typically very weak, sphaleron suppression is insufficient, and such transitions are not viable for baryogenesis. They may, however, lead to interesting gravitational wave signals.

\begin{figure}[h!]
\centering
{\includegraphics[width=0.49\linewidth]{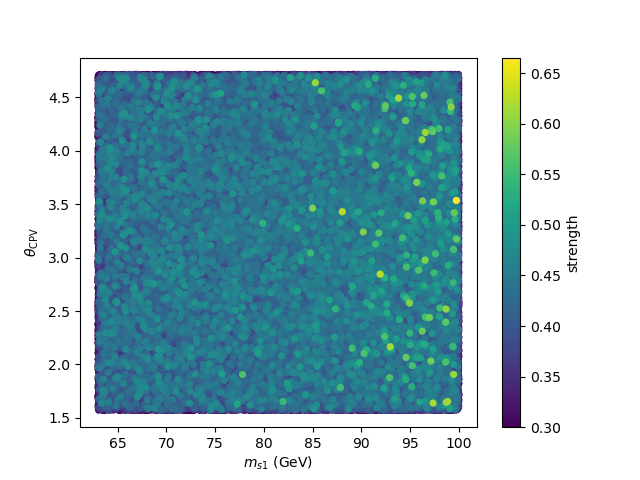}}
{\includegraphics[width=0.49\linewidth]{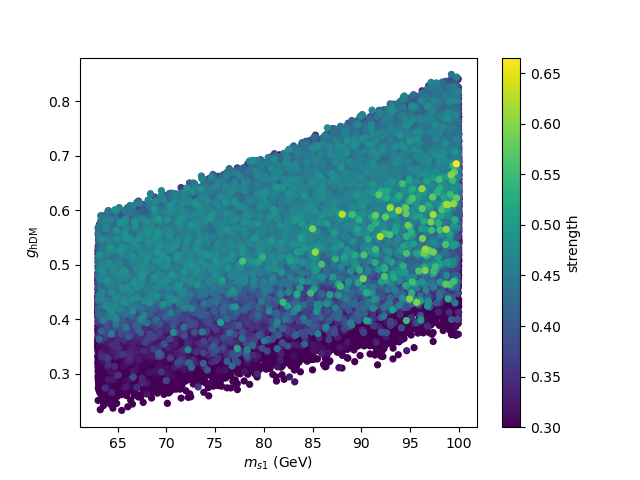}}
{\includegraphics[width=0.49\linewidth]{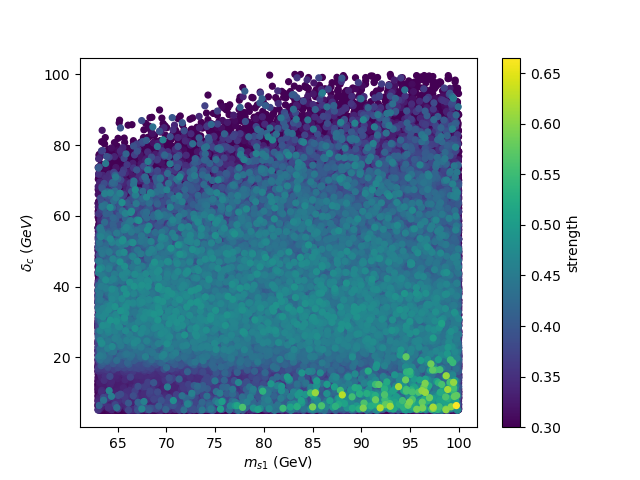}}
{\includegraphics[width=0.49\linewidth]{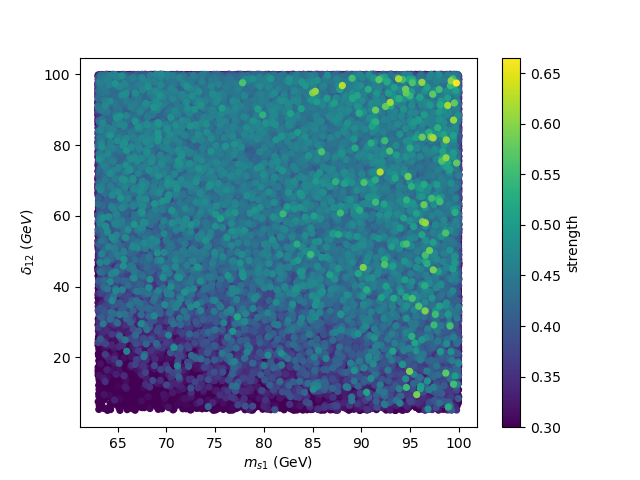}}
\caption{Heat maps of the strength of the $(v_1, v_2, 0) \to (0, 0, v_3)$ transition at one-loop. Step size: $1\,\GeV$.}
\label{fig:2StepScanFigures-1}
\end{figure}

\begin{figure}[h!]
\centering
{\includegraphics[width=0.49\linewidth]{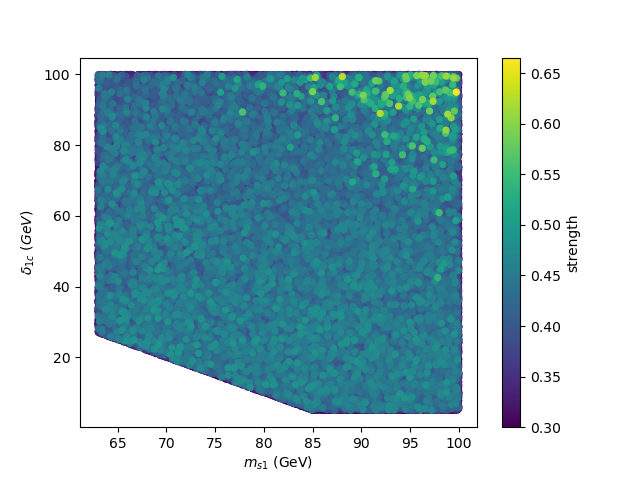}} 
{\includegraphics[width=0.49\linewidth]{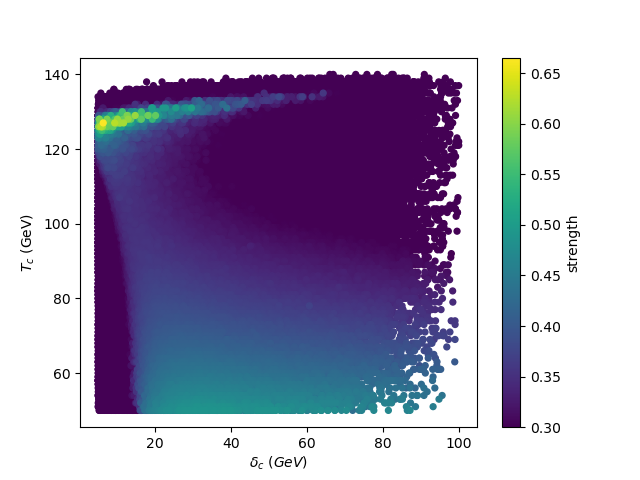}}
{\includegraphics[width=0.49\linewidth]{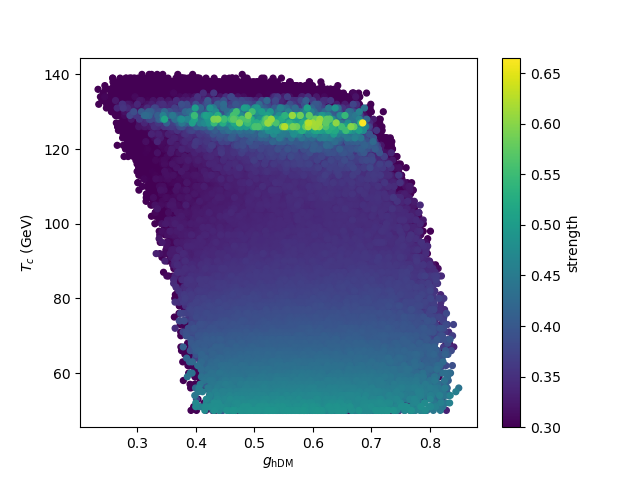}}
{\includegraphics[width=0.49\linewidth]{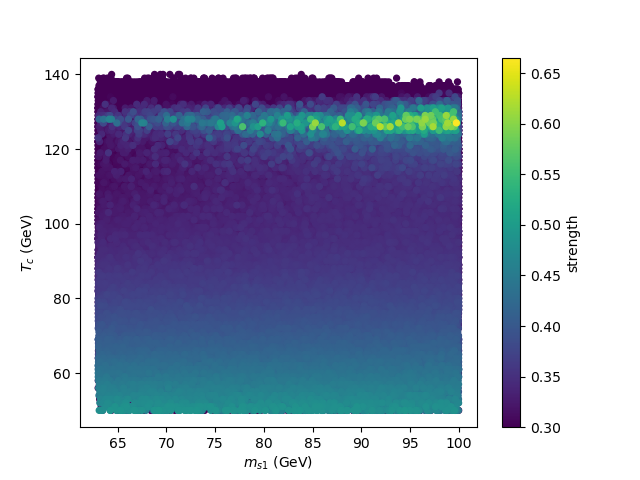}}
\caption{Heat maps of the strength of the $(v_1, v_2, 0) \to (0, 0, v_3)$ transition at one-loop. Step size: $1\,\GeV$.}
\label{fig:2StepScanFigures-2}
\end{figure}

\begin{figure}[h!]
\centering
\includegraphics[angle=-0,width=0.49\linewidth]{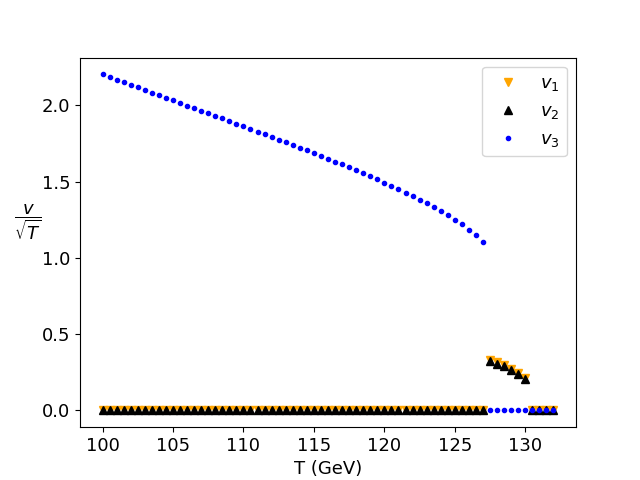}
\includegraphics[angle=-0,width=0.49\linewidth]{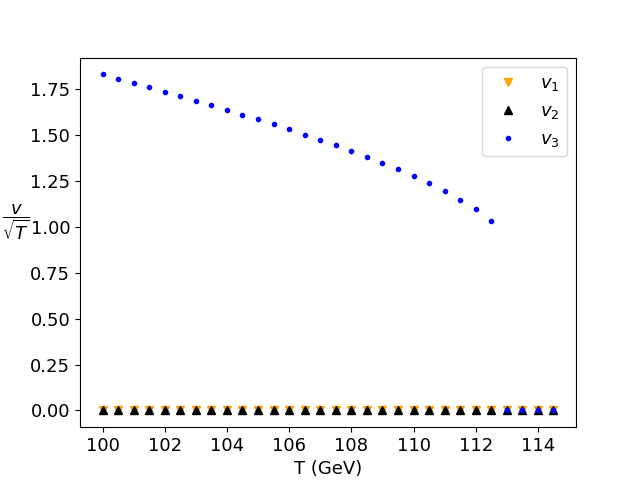}
\caption{Evolution of the vacuum location with temperature for the benchmark in Eq.~\eqref{eq:MaximalCPV}. one-loop (left), two-loop (right).}
\label{fig:VEVtracking}
\end{figure}

Figs.~\ref{fig:2StepScanFigures-1}-\ref{fig:2StepScanFigures-2} show the strength of the second transition. While qualitative behaviour is similar to the one-step case, the range of $T_c$ is broader. As most two-step transitions are weak, perturbative results are unreliable and require confirmation by lattice methods. In the few cases where we performed two-loop calculations, we again find that the strongest two-step transition at one-loop corresponds to a one-step strong first-order phase transition at two-loop level. This is illustrated in Fig.~\ref{fig:VEVtracking} for the benchmark in Eq.~\eqref{eq:MaximalCPV}.

\clearpage
\section{Conclusion and outlook}
\label{sec:conclusions}

We have analysed the electroweak phase transition in a 3-Higgs-doublet model with a CP-violating dark sector, motivated by the simultaneous presence of dark matter and electroweak baryogenesis. Employing high-temperature dimensional reduction and the \texttt{DRalgo} framework, we computed the two-loop thermal effective potential and performed a systematic scan over the parameter space of the model.

Our results show that strong first-order phase transitions are readily achievable in sizeable regions of parameter space. We found that two-loop corrections significantly affect the strength and critical temperature of the transition, with a typical $\sim$27\% reduction in $\Delta v/\sqrt{T_c}$ compared to one-loop estimates. While perturbative predictions overestimate the transition strength, the two-loop calculation provides more accurate guidance for identifying viable benchmarks.

We identified a benchmark point with both a strong first-order phase transition and maximal CP-violating phase, making it a promising candidate for electroweak baryogenesis. Additionally, we observed a subset of scenarios exhibiting two-step transitions, which, though not suitable for baryogenesis due to insufficient sphaleron suppression, may yield interesting gravitational wave signatures.

Our analysis illustrates the importance of going beyond the one-loop approximation in BSM scenarios with extended scalar sectors. The methodology presented here, based on dimensional reduction, symbolic matching, and numerical evaluation, can be readily applied to other multi-scalar models of baryogenesis and dark matter. A dedicated study of more elaborate and exotic vacua the baryon asymmetry generation, including transport dynamics and sphaleron rate calculations, is left for future work.

\subsubsection*{Acknowledgements}
The authors would like to thank Lauri Niemi who was involved in the initial stages of the project. The authors have benefited from many discussions with Kari Rummukainen, David Weir, Oliver Gould and Tuomas Tenkanen.
VK and JTC acknowledge financial support from the Research Ireland Awards Grant 21/PATH-S/9475 (MOREHIGGS) under the SFI-IRC Pathway Program.

\bibliographystyle{JHEP} 
\bibliography{DEWBG.bib}

\end{document}